\documentclass[12pt]{article}
\usepackage{enumitem}
\usepackage{fullpage}
\usepackage[english]{babel}
\usepackage[utf8]{inputenc}
\usepackage[authoryear,comma]{natbib}
\usepackage{amsmath,multibib}
\usepackage{bbm}
\usepackage{bm}
 \usepackage{relsize}
\usepackage{amssymb,amsthm,mathabx,mathtools}
\usepackage{mathrsfs}
\usepackage{graphicx}
\usepackage{url}
\usepackage{caption,subcaption}
\usepackage{booktabs}
\usepackage{multirow}
\usepackage{caption}
\usepackage{epsfig,float}
\usepackage{url} 
\usepackage[usenames,dvipsnames]{color} 
\usepackage{listings}
\usepackage{mathrsfs,bm} 

\usepackage{tikz}
\usepackage{graphicx}

\theoremstyle{plain}
\newtheorem{thm}{Theorem}

\newcommand{\bthm}{\begin{thm}}
\newcommand{\ethm}{\end{thm}}
\newcommand{\bpf}{\begin{proof}}
\newcommand{\epf}{\end{proof}}
\theoremstyle{definition}
\newtheorem{defn}{Definition}
\newtheorem{example}{Example}
\newtheorem{rem}{Remark}
\usepackage[margin=.9in,a4paper]{geometry}
\usepackage[compact]{titlesec}
\renewcommand{\baselinestretch}{1.57}
\numberwithin{equation}{section}
\usepackage{deep-macro}
\usepackage{epigraph}

\usepackage{relsize}
\usepackage[most]{tcolorbox}
\usetikzlibrary{automata, positioning}

\usepackage{bbm}
\usepackage{tabularx}
\newcolumntype{Y}{>{\centering\arraybackslash}X}
\newcolumntype{h}{>{\hsize=.5\hsize\centering\arraybackslash\extracolsep{.1em}}X}
\newcolumntype{q}{>{\hsize=.75\hsize\centering\arraybackslash\extracolsep{.1em}}X}

\usepackage{deep-macro}
\newcommand{\dFnull}{d \hspace{-.08em}\circ \hspace{-.08em}F_0}

\usepackage{cancel}
\usepackage{setspace}

\usepackage{tikz}
\usetikzlibrary{arrows, arrows.meta, calc, positioning, quotes, shapes, shapes.geometric}
\tikzstyle{block} = [rectangle, draw, text centered,    text width=7em, rounded corners, minimum height=1cm]
\tikzstyle{line} = [draw, -latex',line width=.22mm]

\begin{document}
\begin{center} 
{\bf 
{\Large  Density Sharpening: Principles and Applications to Discrete Data Analysis}\\[1.2em]
Deep Mukhopadhyay}\\[.1em]
\texttt{deep@unitedstatalgo.com}\\[.3em]

\end{center}
\vskip.25em
\begin{abstract} 
This article introduces a general statistical modeling principle called ``Density Sharpening'' and applies it to the analysis of discrete count data. The underlying foundation is based on a new theory of nonparametric approximation and smoothing methods for discrete distributions which play a useful role in explaining and uniting a large class of applied statistical methods.  The proposed modeling framework is illustrated using several real applications, from seismology to healthcare to physics.
\end{abstract} 

\noindent\textsc{\textbf{Keywords}}: 
Density sharpening; $\DS(p_0,m)$ distributions; LP-Fourier analysis; Explanatory goodness-of-fit; Jaynes' dice problem; Compressive $\chi^2$; Data-efficient learning.
\vskip1.7em

\renewcommand{\baselinestretch}{-.2}
\setlength{\parskip}{.5ex}
{\small
\setcounter{tocdepth}{2}
\tableofcontents
}
\setlength{\parskip}{1.4ex}
\setstretch{1.58}
\newpage 

\section{Principles of Statistical Model Building}
\begin{quote}
    `\textit{Part of a meaningful quantitative analysis is to look at models and try to figure out their deficiencies and the ways in which they can be improved.}'
\begin{flushright}
\vspace{-.14em}
{\rm ----Nobel Lecture by Lars Peter \cite{hansen2014nobel}} \end{flushright}
\end{quote}
\vspace{-.25em}

Scientific investigation never happens in a vacuum. It builds upon previously accumulated knowledge instead of starting from scratch. Statistical modeling is no exception to this rule. 

\vskip.34em
Suppose we are given $n$ random samples $X_1,\ldots,X_n$ from an unknown discrete distribution $p(x)$. Before we jump into the statistical analysis part, the scientist provided us a hint on what might be an initial believable model for the data: `from my years of experience working in this field, I expect the underlying distribution to be somewhat close to $p_0(x)$.' This information came with a disclaimer: `don't take $p_0(x)$ too seriously as it is only a simplified approximation of reality. Use it with caution and care.'
\vskip.34em

The general problem of statistical learning then aims to address the following questions: Whether the `shape of the data' is consistent with the presumed model-0. If it is not, then what is it? How is it different from $p_0$? Revealing \textit{new} hidden pattern in the data is often the most essential statistical modeling task in science and engineering. Of course, ultimately, the aim is to search for a rich class of sensible models in an automatic and faster manner, by appropriately changing the misspecified $p_0$. Knowing \textit{how} to change the anticipated $p_0$ is the first step towards scientific discovery that allows scientists to re-evaluate alternative theories to explain the data. If we succeed at this, it will provide a mechanism to build ``hybrid'' knowledge-data integrated models, which are far more interpretable than classical fully data-driven nonparametric models.  Full development of these ideas requires a new conceptual framework and mathematical tools. 


\vskip.34em

\textit{Organization}. Section \ref{sec:theory} introduces a new family of nonparametric approximation and smoothing techniques for discrete probability distributions, which is built on the principle of `Density Sharpening.' Section \ref{sec:app} highlights the role of the proposed theoretical framework in the development of statistical methods that is rich enough to include traditional as well as contemporary statistical methods: starting from as simple as one sample Z-test for a proportion to as sophisticated as compressive chi-square, $d$-sharp negative Binomial distribution, universal goodness-of-fit program, relative entropy estimation, Jaynes dice problem, sample-efficient learning of big distributions, etc. The paper ends with a discussion and conclusion Section \ref{sec:diss}.  Additional applications and methodological details are deferred to the Supplementary Appendix to ensure the smooth flow of the main ideas.

\section{Density Sharpening: Model and Mechanism}
\label{sec:theory} 
We describe a method of nonparametric approximation of discrete distribution \textit{by} sharpening the initially assumed $p_0(x)$. The theory is remarkably simple, yet general enough to be vastly applicable in many areas \textit{beyond} density estimation, as described in Section \ref{sec:app}. Here is a bird's eye view of the core mechanism, which is a three-stage process.
\vskip.35em
\texttt{Stage 1}. Model-0 elicitation: The modeler starts a suitable $p_0(x)$ by using his/her experience or subject-matter knowledge. Often a particular parametric form of $p_0(x)$ is selected keeping convenience and simplicity in mind.

\vspace{-.1em}

\texttt{Stage 2}. Exploratory uncertainty analysis: Assess the uncertainty of the presumed model $p_0(x)$, in a way that can explain `why and how' the assumed model-0 (i.e., $p_0$) is inadequate for the data.
\vspace{-.1em}

\texttt{Stage 3}. Coarse-to-Refined density: Incorporate the `learned' uncertainty into $p_0(x)$ to produce an improved model $\hp(x)$ that will eliminate the incompatibility with the data. 
\vskip.65em

The required theory is developed in the next few sections, which heavily relies on the following notation: let $X$ be a discrete variable with probability mass function $p_0(x)$, cumulative distribution function $F_0(x)$, and mid-distribution function $\Fmn(x)=F_0(x) - \frac{1}{2}p_0(x)$. The associated quantile function will be denoted by $Q_0(u)=\inf\{x: F_0(x) \ge u\}$ for $0<u<1$. By $\cL^2(dF_0)$ we mean the set of all square integrable functions with respect to the discrete measure $\dd F_0$, i.e, for a function $\psi \in \cL^2(dF_0)$: $\int |\psi|^2 \dd F_0 := \sum_x |\psi(x)|^2 p_0(x) < \infty$. The inner product of two functions $\psi_1$ and $\psi_2$ in $\cL^2(dF_0)$ will be denoted by $\langle \psi_1, \psi_2 \rangle_{F_0}:=\int \psi_1 \psi_2 \dd F_0$. Expectation with respect to $p_0(x)$ will be abbreviated as $\Ex_0(\psi(X)) :=\int \psi \dd F_0$.

\subsection{Learning by Comparison: $d$-Sharp Density}
We introduce a mechanism for nonparametrically estimating the density of $X_1,\ldots, X_n$ \textit{by comparing and sharpening} the presumed working model $p_0(x)$. \vskip.3em
\begin{defn}[$d$-Sharp Density]
For a discrete random variable $X$, we have the following universal density decomposition:
\beq \label{eq:gd}
p(x)\,=\,p_0(x)\,d\big(F_0(x);F_0,F\big), \eeq
where the $d(u;F_0,F)$ is defined as 
\beq d(u;F_0,F)= \dfrac{p(Q_0(u))}{p_0(Q_0(u))}, ~\,0<u<1.\eeq
The function $d(u;F_0,F)$ is called `comparison density' because it \textit{compares} the assumed $p_0$ with the true $p(x)$ and it integrates to one:
\[\int _0^1 d(u;F_0,F)\dd u \,=\, \int_x d(F_0(x);F_0,F) \dd F_0(x) \,=\,\sum_x \big(p(x)/p_0(x)\big) p_0(x)\,=\, 1. ~~\]
For brevity's sake, we will often 
abbreviate $d(F_0(x);F_0,F)$ as $d_0(x)$ throughout the article.
\end{defn}

\begin{rem}[The philosophy of `\textit{learning by comparison}']
The density representation formula \eqref{eq:gd} describes a way of building a
general $p(x)$ \textit{by comparing it with} the initial $p_0(x)$. The $d$-modulated class of distributions is constructed by amending (instead of abandoning) the starting imprecise model $p_0(x)$.  
\end{rem}

\begin{rem}[$d$-sharp density]
Eq. \eqref{eq:gd} provides a formal statistical mechanism for sharpening the initial vague $p_0(x)$ using the data-guided perturbation function $d_0(x)$. For this reason, we call the improved $p_0(x) \times d_0(x)$ the `$d$-sharp' density. 
\end{rem}
\begin{example}[Earthquake Data]\label{example1}
We are given annual counts of major earthquakes (magnitude 6 and above) for the years 1900-2006. It is available in the R package \texttt{astsa}. Seismic engineers routinely use negative binomial distribution for modeling earthquake frequency \citep{kagan2000prob,kagan2010}. The best fitted negative binomial (NB) with ($\mu=19$ and $\phi=12$) is shown in Fig \ref{fig:earthq}, which we take as our rough initial model $p_0(x)$. From the figure, it is clearly evident that the conventional NB distribution is unable to adequately capture the shape of the earthquake count data. 
\vskip.3em
{\bf Model uncertainty quantification}. For earthquake engineers it is of utmost importance to determine the uncertainty of the assumed NB model \citep{bernreuter1981seismic}. 
The problem of uncertainty quantification is the holy grail of earthquake science due to its importance in estimating risk and in making an accurate forecast of the next big earthquake. 
The comparison density $d(u;F_0,F)$ captures the uncertainty of the assumed NB model. The left plot in Fig \ref{fig:earthq} displays the estimated $\whd(u;F_0,F)$ for this data, which reveals the nature of deficiency of the base NB model. A robust nonparametric method of estimating $\whd$ from data will be discussed in the subsequent sections. But before going into the nonparametric approximation theory, we will spend some time on its interpretation.


\begin{figure}[ ]
\vspace{-1.1em}
  \centering
\includegraphics[width=.46\linewidth,keepaspectratio,trim=1.15cm 1cm 1cm 1cm]{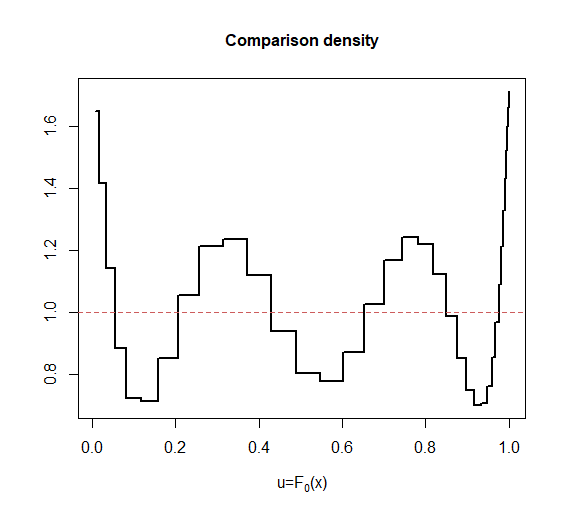}~~~~~~ \includegraphics[width=.46\linewidth,keepaspectratio,trim=1cm 1cm 1.15cm 1cm]{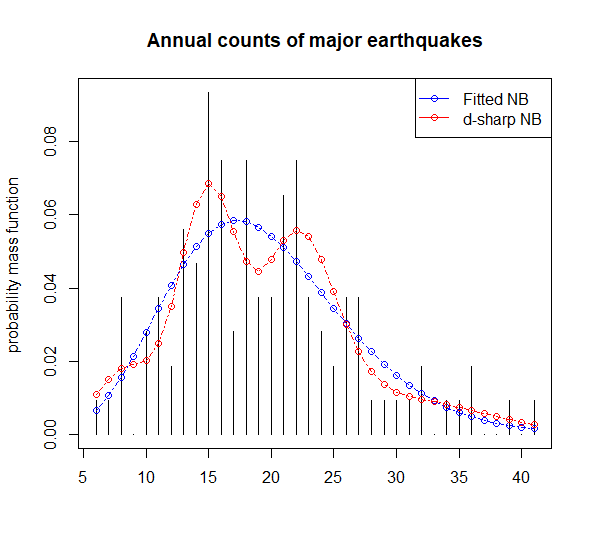}
\vskip.3em
\caption{Modeling the earthquakes distribution. Left: Estimated comparison density $\whd(u;F_0,F)$; Right: The fitted NB distribution and the re-calibrated $d$-sharpened version.}\label{fig:earthq}
\end{figure}


\vskip.2em
{\bf Interpretable Exploratory learning}. In our model \eqref{eq:gd}, $d$ plays the role of a data-driven correction function, measuring the discrepancy between the initial $p_0$ and the unknown $p$. Thus, the non-uniformity of $\whd$ immediately tells us that there's something more in the data than what was expected in light of $p_0(x)$.  In fact, the shape of $\whd$ reveals the nature of the most prominent deviations between the data and the presumed $p_0(x)$---which, in this case, are bimodality and presence of heavier-tail than anticipated NB distribution.
\vspace{-.3em}
 \end{example}
\begin{rem}[Role of $d$]
The comparison density $d$ performs dual functions: (i) its graph acts as an exploratory diagnostic tool that exposes the unexpected,
forcing decision makers (e.g., legislators, natural security agencies, local administrations) to think outside the box: what might have caused this bimodality? how can we repair the old seismic hazard forecast model so that it incorporates this new information? etc. (ii) it provides a formal process of transforming and revising an initially misspecified model into a useful one. The red curve in the right panel is obtained by multiplying (perturbing) the NB pmf with the estimated comparison density, obeying the density representation formula Eq. \eqref{eq:gd}.
\end{rem}







\subsection{LP-Fourier Analysis}
\label{sec:LPFA}
The task is to nonparametrically approximate $d(F_0(x);F_0,F)$ to be able to apply the density sharpening equation \eqref{eq:gd}. We approximate $\dFnull(x) \in \cL^2({dF_0})$ by projecting it into a space of polynomials of $F_0(x)$ that are orthonormal with respect to the base measure$\dd F_0$. How to construct such a system of polynomials in a completely automatic and robust manner for \textit{any} given $p_0(x)$? In the section that follows, we discuss a universal construction. 
\subsubsection{Discrete LP-Basis}
We describe a general theory of constructing LP-polynomials---a new class of robust polynomials $\{T_j(x;F_0)\}_{j\ge 1}$ that are a function of $F_0(x)$ (not raw $x$) and are orthonormal with respect to user-specified discrete distribution $p_0(x)$.

\begin{figure}[ ]
  \centering
\includegraphics[width=.45\linewidth,keepaspectratio,trim=1cm 1cm 1cm 1cm]{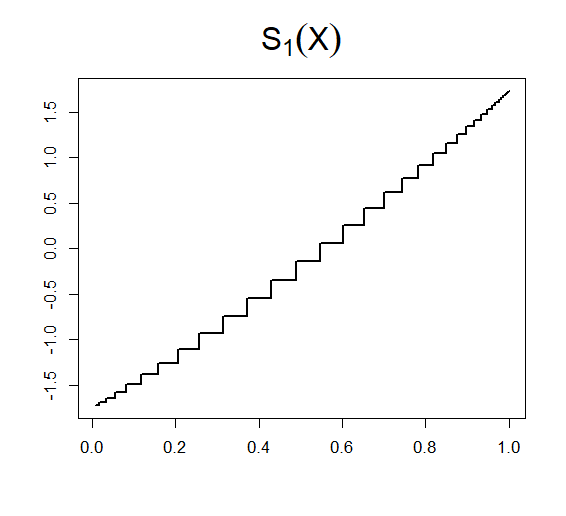}~~~~~
\includegraphics[width=.45\linewidth,keepaspectratio,trim=1cm 1cm 1cm 1cm]{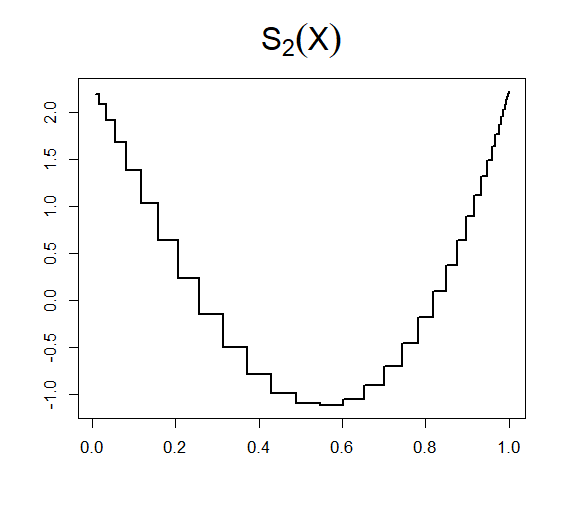}\\[2em]
\includegraphics[width=.45\linewidth,keepaspectratio,trim=1cm 1cm 1cm 1cm]{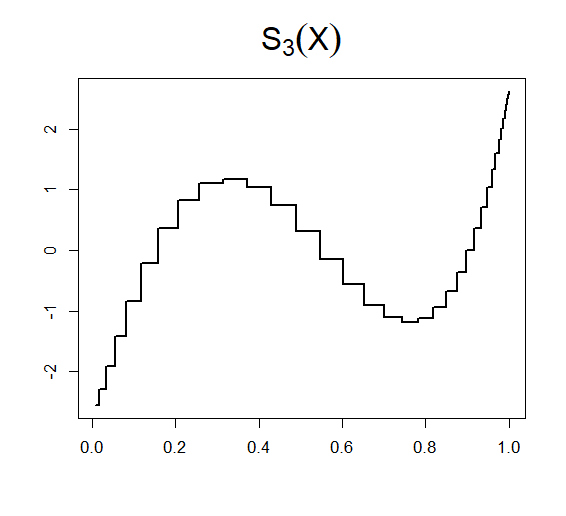}~~~~~
\includegraphics[width=.45\linewidth,keepaspectratio,trim=1cm 1cm 1cm 1cm]{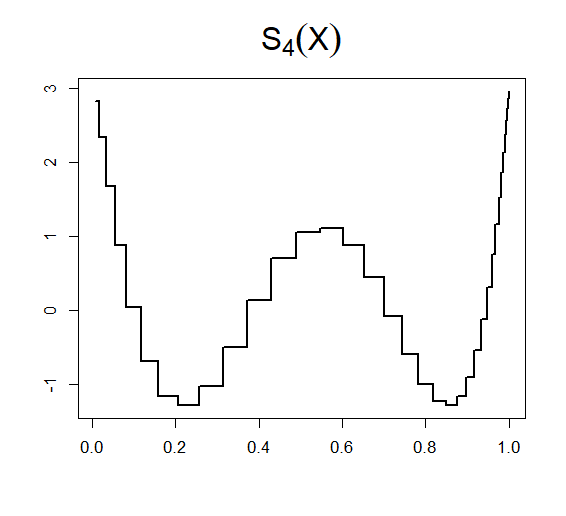}
\vskip1.5em
\caption{Shape of the top four LP-orthonormal basis functions $\{S_j(u;F_0)\}_{1\le j \le 4}$ for $p_0(x) :={\rm NB}(x;\mu=19,\phi=12)$ for the earthquake data; where recall $S_j\hspace{-.084em}\circ \hspace{-.084em} F_0\,(x)=T_j(x;F_0)$. Notice of the global nonlinearity (linear, quadratic, cubic and so on) and the local staircase-like (piecewise-constant unequal-length segments) shape of these specially-designed polynomials. They are, by construction, orthonormal with respect to the chosen measure $p_0(x)$, here ${\rm NB}(x;\mu=19,\phi=12)$.}\label{fig:earthT}
\end{figure}

\texttt{Step 1:} Define the first-order LP-basis function as \textit{standardized} mid-distribution transform:
\beq \label{eq:T1} 
T_1(x;F_0) \,=\,\dfrac{\sqrt{12} \big[\Fmn(x) - 0.5\big]}{\sqrt{1-\sum_x p_0^3(x)}}. \eeq
Verify that $\Ex_0[T_1(X;F_0)]=0$ and $\Ex_0[|T_1(X;F_0)|^2]=1$, since $\Ex[\Fmn(X)]=1/2$ and $\Var[\Fmn(X)]=\sqrt{(1-\sum_x p_0^3(x))}/12$.
\vskip.25em

~~\texttt{Step 2:} Apply a \emph{weighted} Gram-Schmidt procedure on $\{T_1^2,\ldots  T_1^{k-1}\}$ to construct a higher-order LP orthogonal system $T_j(x;F_0)$ with respect to measure$\dd F_0$ 
\[\sum\nolimits_x p_0(x) T_j(x;F_0)=0;~~\,\sum\nolimits_x p_0(x) T_j(x;F_0)T_k(x;F_0)=\delta_{jk}, ~~1<j,k<M \]
where $\delta_{jk}$ is the Kronecker delta function and the highest-degree of the LP-polynomials $M$ is always less than the support size of the discrete $p_0$. For example, if $X$ is binary, one can construct at most $2-1=1$ LP-basis function; see Section \ref{sec:bin}.
\vskip.3em
Fig \ref{fig:earthT} shows the top four LP-basis functions for the earthquake data with $p_0$ as ${\rm NB}(x;\mu=19,\phi=12)$. Here, we have displayed them in a unit interval as a function of $u=F_0(x)$, denoted by $S_j(u;F_0) := T_j(Q_0(u); F_0), 0<u<1$. Notice the typical shape of these custom-constructed discrete orthonormal polynomials: globally nonlinear (linear, quadratic, cubic, and so on) and locally piecewise-constant with unequal step size. 
\begin{rem}[Role of LP-coordinate system in unification of statistical methods]
LP-bases play a unique role in statistical modeling---they provide an efficient coordinate (data-representation) system that is fundamental to developing unified statistical algorithms.
\vspace{-.65em}
\end{rem}


\subsection{The {\boldmath$\DS(p_0,m)$}  Model} \label{sec:dstheory}
\begin{defn}[LP-canonical Expansion]
Expand comparison density in the LP-orthogonal series
\beq \label{cdm}
d(F_0(x);F_0,F)\,=\,1+\sum_j \LP[j;F_0,F] \,T_j(x;F_0),
\eeq
where the $j$th LP-Fourier coefficient satisfies the following identity:
\beq \label{eq:dlp}
\LP[j;F_0,F]= \big\langle \dFnull, T_j \big \rangle_{F_0}.\eeq
A change-of-basis perspective: The conventional way to represent a discrete distribution is through indicator basis (histogram representation):
\beq
\eta_j(i)\,=\, \ind \big\{ X_i \in [x_j, x_{j+1} )  \big\},~~ {\rm for}\, j=1,2,\ldots, r~
\eeq
where $r$ is the domain size (number of unique values) of the empirical distribution $\tp(x)$. In \eqref{cdm}, we have performed a ``change of basis'' from the amorphous indicator-basis to a more structured LP-basis $\{T_j(x;F_0)\}$, where the expansion coefficients $\LP[j;F_0,F]$ act as the coordinates of $p(x)$ \textit{relative to} assumed $p_0(x)$:
\[\big[ F \big]_{F_0} := \Big(\LP[1;F_0,F], \ldots, \LP[m;F_0,F]\Big),~~1\le m < r.\]
For that reason, one may call these coefficients the discrete LP-Fourier Transform (LPT) of $p(x)$ relative to $p_0(x)$.
\end{defn}
\begin{defn}
$\DS(p_0,m)$ denotes a class of distributions with the following representation:
\beq  \label{DSm}
p(x)\,=\,p_0(x)\Big[ 1\,+\, \sum_{j=1}^m \LP[j;F_0,F]\, T_j(x;F_0)\Big],
\eeq
obtained by replacing \eqref{cdm} into \eqref{eq:gd}. Here $\DS(p_0,m)$ stands for {\bf D}ensity-{\bf S}harpening of $p_0(x)$ using $m$-term LP-series approximated $d_0(x)$. $\DS(p_0,m)$ is a class of nonparametrically-designed parametric models that are flexible enough to capture various \emph{shapes of discrete} $p(x)$, like multi-modality, excess-variation, long-tailed, and sharp peaks.
\vspace{-.65em}
\end{defn}
To estimate the unknown coefficients $\LP[j;F_0,F]$ of the $\DS(p_0,m)$ model, note the following important identity: 
\vspace{-.35em}
\bea 
\label{eq:lpeq}
\LP[j;F_0,F]&=& \int d(F_0(x);F_0,F) T_j(x;F_0) \dd F_0(x) \nonumber \\
&=& \int T_j(x;F_0) \dd F(x) \nonumber \\
&=& \Ex_F\big[  T_j(X;F_0) \big].
\eea
\vspace{-.4em}
This immediately leads to the following ``weighted mean'' estimator:
\beq \label{eq:eestlp}
\tLP_j\,:= \LP[j;F_0;\widetilde F]\,=\,  \Ex_{\wtF}\big[T_j(X;F_0)\big]\,=\,\sum_x \tp(x) T_j(x;F_0),~~~~
\eeq
Using standard empirical process theory \citep{csorgHo1983quantile,parzen1998statistical} one can show that the limiting distribution of sample LP-statistic is i.i.d $\cN(0,n^{-1/2})$, under the null hypothesis $H_0:p=p_0$. Thus one can quickly obtain a sparse estimated $\DS(p_0,m)$ model by retaining only the `significant' LP-coefficients, which are greater than $2/\sqrt{n}$.

{\bf Earthquake Data Example}. The first $m=10$ estimated $\tLP_j$ are shown in Fig. \ref{fig:earthlp} of the appendix, which indicates that the only interesting non-zero LP-coefficient is $\tLP_6$. The explicit form of the estimated $\DS({\rm NB},m=6)$ model for the earthquake data is given by:
\beq \label{eq:dsequ} 
\hp(x) = p_0(x)\big[ 1 + 0.20 T_6(x;F_0)  \big],\eeq
where $p_0={\rm NB}(x;\mu=19, \phi=12)$. The resulting $\hp(x)$ is plotted as a red curve in Fig. \ref{fig:earthq}. 
\subsection{LP-Maximum Entropy Analysis}
\label{sec:lpmaxent}
To ensure non-negativity, we expand $\log d$ (instead of $d$ as we have done in Eq. \eqref{cdm}) in LP-Fourier series, which results in the following exponential model:
\beq \label{eq:dexp}
d_{\teb}(u;F_0,F)\,=\,\exp\Big \{ \sum_{j\ge 1} \te_j S_j(u;F_0)\,-\, \Psi(\teb)\Big \},~~0<u<1
\eeq
where $\Psi(\teb)=\log \int_0^1 \exp\{ \sum_j \te_j S_j(u;F_0)\}\dd u.$
This model is also called the maximum-entropy (maxent) comparison density model because it maximizes the entropy $-\int d_{\teb} \log d_{\teb}$ (flattest possible; thus promotes smoothness) under the following LP-moment constraints:
\beq \label{eq:cons}
\Ex_{\teb}[S_j(U;F_0)]\,=\,\LP[j;F_0,\wtF],~~(j=1,2\ldots).
\eeq
LP-moments $\LP[j;F_0,\wtF]$ are `compressed measurements' (linear combinations of observed data; verify from \eqref{eq:eestlp}), which are sufficient statistics for the comparison density $d_{\teb}$. 
\vskip.4em

\begin{defn}[Maxent $\DS(p_0,m)$ model]
Replacing \eqref{eq:dexp} into \eqref{eq:gd}, we have the following maxent  $\DS(p_0,m)$ model
\beq \label{eq:maxentds}
p(x)\,=\,p_0(x) \exp\Big \{ \sum_{j\ge 1} \te_j T_j(x;F_0)\,-\, \Psi(\teb)\Big\},
\vspace{-.5em}
\eeq
To estimate a sparse maxent comparison density model, we carry out the optimization routine by choosing only the `significant' LP-moments in \eqref{eq:cons}.
\end{defn}
\vskip.1em

{\bf Earthquake Data Example}. 
The estimated maxent $\DS(p_0,m)$ model for the earthquake distribution is given by
\beq \label{eq:xnbequake}
\hhp(x) = p_0(x)\exp\big \{  0.195 T_6(x;F_0) - 0.02\big \},\eeq
whose shape is almost indistinguishable from the LP-Fourier estimated p.m.f. \eqref{eq:dsequ}.







\section{Applications in Statistical Modelling} \label{sec:app}
We describe how the general principle of `density sharpening' acts as a unified framework for the analysis of discrete data with a wide variety of applications, ranging from basic introductory methods to more advanced statistical modeling techniques.
\subsection{One-sample Test of Proportion} \label{sec:bin}
Given $n$ samples from a binary $X$, the one-sample proportion test is concerned with testing whether the population proportion $p$ is equal to the hypothesized proportion $p_0$. We approach this problem by reformulating it in our mathematical notation:

Step 1. We start with the $\DS(p_0,m=1)$ model 
\beq \label{eq:dsbin}
p(x)=p_0(x) \Big\{1+\LP[1;p_0,p] T_1(x;F_0)\Big\},~~x=0,1.\eeq
where the null model $p_0(x)= xp_0 + (1-x) (1-p_0),$ for $x=0,1$.

Step 2. We rewrite the original hypothesis $H_0:p=p_0$ in terms of LP-parameter as $H'_0:\LP[1;p_0,p]=0$.

Step 3. We derive an explicit formula for $\LP[1;p_0,\tp]$. It's a two step process: First, we need the analytic expression of the LP-basis $T_1(x;p_0)$
\beq T_1(x;p_0) = \left\{ \begin{array}{rl}
 -\dfrac{p_0}{\sqrt{p_0(1-p_0)}} &\mbox{for $x=0$} \\
 \dfrac{1-p_0}{\sqrt{p_0(1-p_0)}} &\mbox{for $x=1$.}
       \end{array} \right. \eeq
We then apply formula \eqref{eq:eestlp} to deduce:
\beq \label{lp1bin} \LP(1;p_0;\tp) = (1-\tp)T_1(0;p_0) + \tp T_1(1;p_0) = \dfrac{\widetilde p - p_0}{\sqrt{p_0(1-p_0)}}.~~~\eeq
Step 4. A remarkable fact is that the test based on \eqref{lp1bin} exactly matches with the classical Z-test, whose null distribution is: $\sqrt{n} \LP[1;p_0,\tp] \sim \cN(0,1)$ as $\nti$. This shows how the LP-theoretical device provides a transparent \textit{first-principle derivation} of the one-sample proportion test, by shedding light on its genesis.


\subsection{Expandable Negative Binomial Distribution}\label{sec:dnb}
We will focus now on one important special case of maxent $\DS(p_0,m)$ family of distributions \eqref{eq:maxentds}, where the base measure $p_0(x)$ is taken to be a negative Binomial (NB) distribution. 
\beq \label{eq:xnb}
p(x) \,= \,\binom{x + \phi - 1}{x} \,
\left( \frac{\mu}{\mu+\phi} \right)^{\!x} \, \left(
\frac{\phi}{\mu+\phi} \right)^{\!\phi} \! \exp\Big \{ \sum_{j\ge 1} \te_j T_j(x;F_0)\,-\, \Psi(\teb)\Big \},~~x \in \mathbb{N},~~\eeq
note that the basis functions $\{T_j(x;F_0)\}_{j\ge 1}$ are specially-designed LP-orthonormal polynomials associated with the base measure $p_0 = {\rm NB}(\phi,\mu)$. We call \eqref{eq:xnb} the $m$th-order expandable NB distributions, denoted by \texttt{XNB}(m).
A few practical advantages of \texttt{XNB}(m) distributions are: the computational ease they afford for estimating parameters; their compactly parameterizable yet shape-flexible nature; and, finally, their ability to provide explanatory insights into how $p(x)$ is different from the standard NB distribution. Due to their simplicity and flexibility, they have the potential to be a `default choice' for modeling count data. We have already seen an example of $\texttt{XNB}$-distribution in \eqref{eq:xnbequake} in the context of modeling the earthquake distribution. We now turn our attention to two further real-data examples.







\begin{figure}[ ]
  \centering
\includegraphics[width=.48\linewidth,keepaspectratio,trim=1cm 1cm 1cm 1cm]{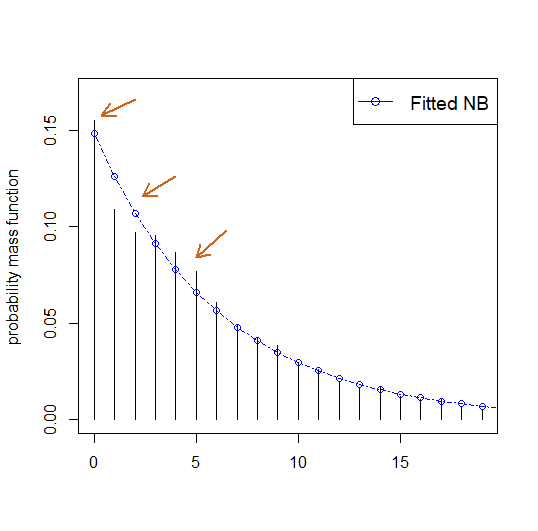}~~~~~
\includegraphics[width=.48\linewidth,keepaspectratio,trim=1cm 1cm 1cm 1cm]{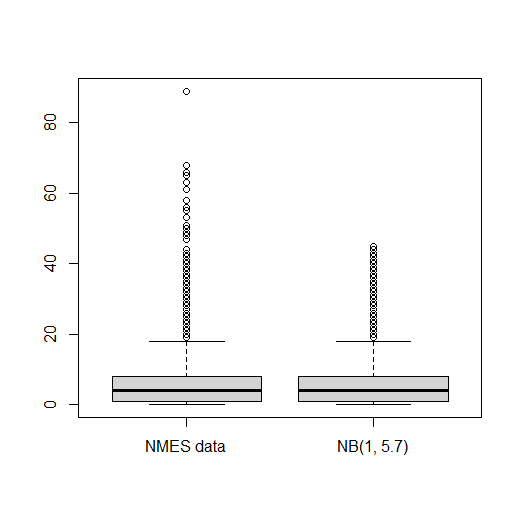}\\[2em]
\includegraphics[width=.48\linewidth,keepaspectratio,trim=1cm 1cm 1cm 1cm]{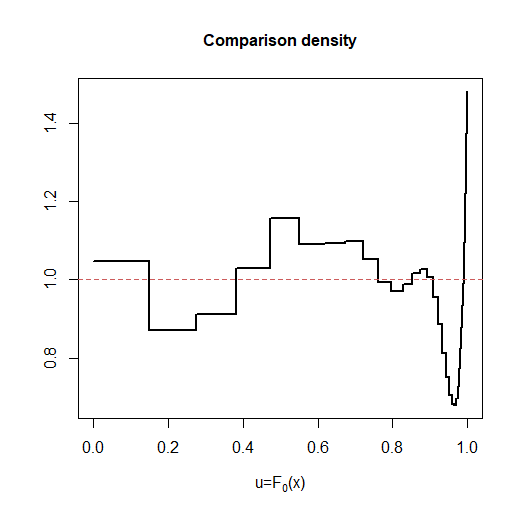}~~~~~~~
\includegraphics[width=.48\linewidth,keepaspectratio,trim=1cm 1cm 1cm 1cm]{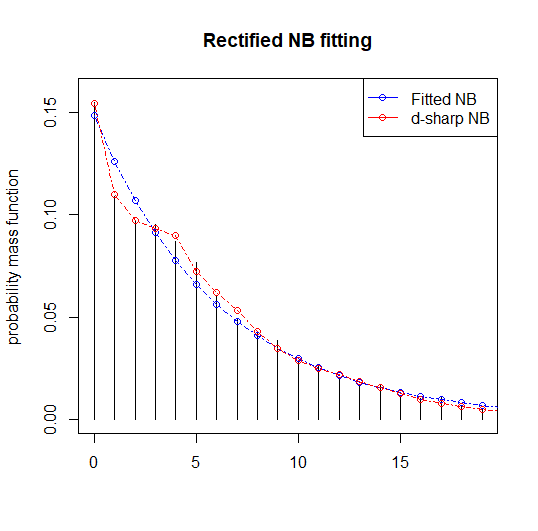}
\vskip1.5em
\caption{Nonparametric modeling of NMES 1988 data. Top left: For clarity, we only show the data over the domain $[0,20]$. Otherwise, the ultra long-tailedness of the data will make it harder to see what's going on in the interesting zones; 
the arrows point out some specific aspects of the data that were missed by NB distribution. Top right: compares two boxplots---one is the observed data and the other is the simulated data from ${\rm NB}(1,5.7)$, which captures the difference in the tail behavior. The tail of NB has to be stretched to match the shape of the data. Bottom left: $\whd$ captures the `unexplained shape of the data,' which tells us \textit{how} to change the initial model-0 to fit the data. Bottom right: The red curve is the rectified \texttt{XNB} probability distribution.}\label{fig:NMES}
\end{figure}

\begin{example}[NMES 1988 Data]
This is a part of the US National Medical Expenditure Survey (NMES) conducted in 1987 and 1988. It is available in the R-package \texttt{AER}. We have $n=4,406$ observations of a discrete random variable $X$ which denotes how many times an individual, aged 66 and covered by Medicare, visited physician's office. As displayed in the Fig. \ref{fig:NMES} boxplot, the distribution has a large support size (varies between $0$ and $89$), with some regions being extremely data-sparse. 

\vskip.35em

The blue curve in the top left plot shows the ${\rm NB}(\hat \phi=1, \hat \mu=5.7)$, where the parameters are maximum-likelihood estimates. Next, we estimate the LP-maxent $\whd_{\teb}$, using the theory of Sec. \ref{sec:lpmaxent}. At this point, it is strongly advisable to pay attention to the shape $\whd_{\teb}$. Why? Because, it efficiently extracts and exposes `unanticipated' aspects in the data that cannot be explained by the initial NB distribution. The bottom-left Fig. \ref{fig:NMES} immediately reveals a few things: (i) NB underestimates the probability at $x=0$; (ii) it overestimates the probability mass around $x=2$ and $3$; (iii) there seems to be an excess probability mass (`bump' structure) around $x=4$; (iv) NB clearly has a shorter tail than what we see in the data---this can be seen from the sharply increasing right tail of the comparison density. To better understand the tail-behavior, we have simulated $n$ samples from ${\rm NB}(1,5.7)$ and contrasted the two boxplots in the top-right panel, which strongly indicates the long-tailedness of $p(x)$ relative to NB distribution.  Any reader will agree that without the help of $\whd_{\teb}$, even experienced eyes could have easily missed these subtle patterns. Finally, multiply $\whd_{\teb}$ by the ${\rm NB}(1,5.7)$, following eq. \eqref{eq:xnb}, to get the estimated \texttt{XNB} distribution---the red p.m.f curve, shown in the bottom-right panel of Fig. \ref{fig:NMES}.
\end{example}

\begin{example}[Computer Breaks Data] \label{ex:comp}
We are given the number of times a DEC-20 computer broke down at Open University in each of $n=128$ consecutive weeks of operation, starting in late 1983. The data shows positive skewness with a  slightly longer tail; see Fig. \ref{fig:comp} in the appendix. The mechanics of \texttt{XNB} modeling proceed as follows: (i) We start by estimating the parameters of $p_0(x)$, which in this case are MLE-fitted (one can use any other method of estimation) ${\rm NB}(\hat \phi=1.7, \hat \mu=4)$. (ii) The next step is estimation of $\whd_{\teb}$, which in this case is just the uniform distribution---none of the LP-maxent parameters were large enough to be selected. This is depicted in the left panel of supplementary Fig. \ref{fig:comp}. This graphical diagnostic indicates that the initial $p_0$ fits the data satisfactorily; no repairing is necessary. (iii) Accordingly, our `density sharpening' principle returns the $\texttt{XNB}(m=0)$  as the final model, which is simply the starting parametric model ${\rm NB}(\hat \phi=1.7, \hat \mu=4)$.

It is interesting to contrast our finding with \cite{saulo2020family}, where the authors fit a highly specialized nonparametric discrete distribution to this data. The beauty of our approach is that it performs nonparametric correction (through $d$) only when it is warranted. When the reality is already simple,  we don't complicate it unnecessarily.
\end{example}
\vspace{-.3em}

\subsection{{\boldmath$\chi^2$} and Compressive-{\boldmath$\chi^2$}}
Given a random sample of size $n$ from the unknown population distribution $p(x)$, chi-square goodness-of-fit statistic between the sample probabilities $\tp(x)$ and the expected $p_0(x)$ can be re-written as follows:
\beq \dfrac{\chi^2}{n} = \sum_x \dfrac{\big( \tp(x) - p_0(x)  \big)^2}{p_0(x)} = \sum_x p_0(x) \big[ \tp(x)/p_o(x)\,-1 \big]^2= \int_0^1 \big[  d(u;p_0,\tp) - 1\big]^2\dd u,\eeq
By applying Parseval's identity on the LP-Fourier expansion of $d$, we have the following important equality:
\beq \label{eq:LPchisq}
\dfrac{\chi^2}{n}\, =\, \sum_{j=1}^{r-1}\Big|  \LP[j;p_0,\tp] \Big|^2\defeq\,\LP(p_0 \| \tp),
\eeq
where $r$ is the number of unique values in our sample $X_1,\ldots,X_n$. This shows that chi-square information statistic is a ``saturated'' raw-nonparametric measure with $r-1$ components.
\begin{example}[The Gambler's Die] \label{ex:gam}
A gambler rolls a die $n=60$ times and gets the following observed counts:
\vskip.25em
\begin{table}[h]
\centering
\caption{The observed frequencies}
\renewcommand{\tabcolsep}{.3cm}
\renewcommand{\arraystretch}{1.2}
\begin{tabular}{lcccccc}
\hline
Number on die & $1$ & $2$  & $3$ & $4$ & $5$ & $6$\\
Observed $\tp$  & 4/60 & 6/60 &17/60 &16/60 & 8/60 &9/60\\
Hypothesized $p_0$ & 1/6 & 1/6 & 1/6 & 1/6 & 1/6 & 1/6\\
  \hline
\end{tabular}
\end{table} \label{tab:gam}
The gambler wishes to determine whether the die is fair. If it is fair, we would expect 
the outcomes of $1$ to $6$ are equally likely, with probability $1/6$. Pearsonian chi-square and a full-rank LP-analysis, both lead to the same answer:
\[
\chi_{{\rm obs}}^2\, =\, n \times \sum_{j=1}^{6-1}\Big|  \LP[j;p_0,\tp] \Big|^2=14.2, ~\,\text{with degrees of freedom $5$}
\]
with the resulting  $p$-value $0.0143$. Note that 
the sum of squares of the $6-1=5$ LP-Fourier coefficients ``exactly'' reproduces (numerically) the Pearson's chisquare statistic! This further verifies the mathematical fact elucidated in eq. \eqref{eq:LPchisq}. Conclusion:  The die is loaded at 5\% significance level. 

\begin{figure}[ ]
  \centering
\includegraphics[width=.46\linewidth,keepaspectratio,trim=2cm 1.5cm 1cm .55cm]{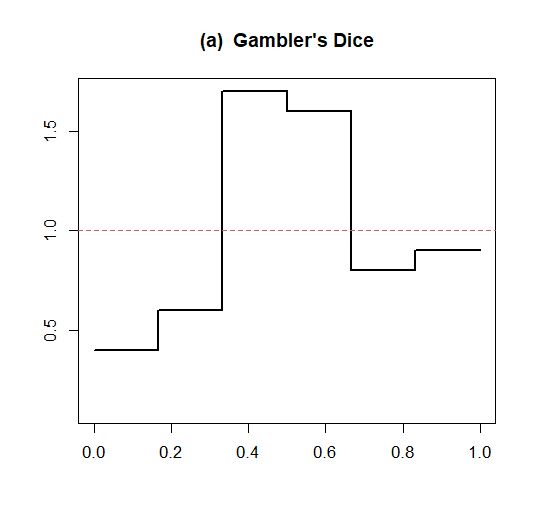}~~~
\includegraphics[width=.46\linewidth,keepaspectratio,trim=1cm 1.5cm 2cm .55cm]{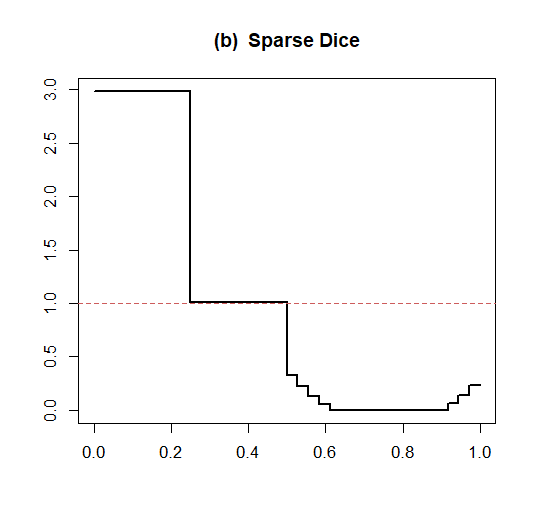}
\caption{The estimated $\whd(u;F_0,F)$ for examples \ref{ex:gam} and \ref{ex:SD}. It helps to identify `where' the interesting differences between data ($\tp$) and hypothesized model ($p_0$) lie.}\label{fig:gamdie}
\end{figure}
\textit{Exploratory insight}. Here we want to go beyond classical confirmatory analysis,  with the goal to understand \textit{how the die is loaded}. The answer is hidden in the shape of the comparison density $\whd$. Fig. \ref{fig:gamdie}(a) firmly suggests that the die was loaded heavily in the middle ---especially on the sides $3$ and $4$, where it landed most frequently. 
\end{example}
 
\begin{example}[Sparse Dice problem]
\label{ex:SD}
It is an example of sparse count data with many groups. Imagine a $k=20$ dimensional dice is rolled $n=20$ times:
\begin{itemize}[itemsep=1pt,topsep=1pt]
  \item The hypothesized model:~~ \,$p_0=(1/4,1/4, 1/36,\ldots, 1/36)$
  \item The observed probabilities:~~$\tp=(3/4, 1/4,0,\ldots,0)$.
\end{itemize}
We would like to know whether the  postulated model $p_0(x)$ actually reflects the data or not. If it does not, then we want to know how the hypothesized model differs from the observed probabilities.  

Pearsonian chi-square\footnote{R-function \texttt{chisq.test}() generates the message that ``Chi-squared approximation may be incorrect!.''} yields value $\chi_{{\rm obs}}^2 =30$, with degrees of freedom $19$ and  $p$-value $0.052$. Conclusion: there is no evidence of discrepancy at 5\% level, even though there is a glaring difference between $\tp(1)=3/4$ and $p_0(1)=1/4$. The legacy $\chi^2$ loses its power because of `inflated degrees of freedom' for large sparse problems: Large value of $k$ increases the critical value $\chi^2_{\al;k-1}$, making it harder to detect `small' but important changes.

\vskip.4em
LP-Analysis of the sparse dice problem: (i) Construct the discrete LP-polynomials $\{T_j(x;p_0)\}$ that are specially-designed for the given $p_0(x)$. Appendix Fig. \ref{fig:SD:basis} displays the shape of those basis functions. (ii) Compute the LP-Fourier coefficients $\LP[j;p_0,\,\tp]$ by $\sum_x \tp(x) T_j(x;p_0)$. The Appx. figure \ref{fig:sdice} identifies the first two LP-parameters as significant components. (iii) We now compute the compressive-$\chi^2$ based on these interesting components:
\beq 
\LP(p_0 \| \tp) \,=\,\big|\LP[1;p_0,\,\tp]\big|^2 \,+\, \big|\LP[2;p_0,\,\tp]\big|^2=1.49,~\,~\text{with degrees of freedom $2$}.
\eeq
and  $p$-value $3.4\times 10^{-7}$. Also noteworthy is the fact that compressive LP-chisquare is numerically almost same as the raw $\chi^2$:
\[
\vspace{-1em}
\chi^2_{{\rm obs}} \,=\, 30 ~\approx~ n \times \LP(p_0 \| \tp) = 29.8.\]
\end{example}

\begin{rem}[Auto-adaptability]
LP-goodness-of-fit shows an impressive adaptability property: under the usual scenario (like in example \ref{ex:gam}) it reduces to the classical $\chi^2$ analysis, and for large-sparse problems, it automatically produces a chi-square statistic with the fewest possible degrees of freedom, which boosts its power. Our reformulation (in terms of modern LP-nonparametric language) allowed a better way of doing chi-square goodness-of-fit analysis that applies to a much broader class of applied statistics problems. In John Tukey's (1954) words: ``Do we need to find new techniques, or to use old ones better?''
\end{rem}

\begin{rem}[Ungrouped case]
What if we have \textit{ungrouped} data: given a random sample of counts $X_1,\ldots,X_n$, check (confirm) whether the data  is compatible with the hypothesised $p_0(x)$, i.e., to test the hypothesis $H_0:p=p_0$. One way to approach this problem is to forcefully group the data points into different categories and then apply Pearson's $\chi^2$ test on it. This is (almost) always a bad strategy, since grouping leaks information. Moreover, the arbitrariness involved in choosing the groups makes it an even less attractive option. However, our LP-divergence measure $\LP(p_0 \| \tp)$ \eqref{eq:LPchisq} can be applied to ungrouped data, with no trouble. The next section expands on this.
\end{rem}


\subsection{Explanatory Goodness-of-Fit}
\label{sec:Xgof}
What is an explanatory goodness-of-fit? Why do we need it? This is perhaps best answered by quoting the views of John Tukey:
\begin{quote}
\vspace{-.4em}
\textit{``What are we trying to do with goodness of fit tests? (Surely not to test whether the models fits exactly, since we know that no model fits exactly!) What then? How should we express the answer of such test?}''
\begin{flushright}
\vspace{-1.35em}
{\rm ---John \cite{tukey1954}}\end{flushright}
\end{quote}
\vspace{-.4em}
To satisfactorily answer these questions we need to design a GOF-procedure that is \textit{simultaneously} confirmatory and exploratory in nature: 
\vskip.3em
~~$\bullet$ On the confirmatory side, it aims to develop a universal GOF statistic that is easy to use and fully automated for \textit{any} user-specified discrete $p_0(x)$. One such universal GOF measure is $\LP(p_0 \| \tp)$, defined as
\beq \label{eq:ugof}
\LP(p_0 \| \tp)\,=\,\int_0^1 \big(d(u;p_0,\tp) - 1\big)^2 \dd u = \sum_j \Big|  \LP[j;p_0,\tp] \Big|^2= \sum_j \Big|\sum_x \tp(x) T_j(x;F_0)\Big|^2,~~
\eeq
where the index $j$ runs over the significant components. See Appx. \ref{app:gof} for more details.

\vskip.3em
~~$\bullet$ On the exploratory side, the graphical visualization of comparison density $d(u;p_0,\tp)$ provides explanations as to \textit{why} $p_0(x)$ is inadequate for the data (if so) and \textit{how} to rectify it to reduce its incompatibility with the data. This has special significance for data-driven discovery and hypothesis generation. In particular, the non-zero LP-coefficients indicate the ``main sources of discrepancies.'' In the following, we illustrate this method using three real data examples, each of which contains a different degree of lack-of-fit.


\begin{example}[Spiegel Family Data] \label{ex:spiegel}
\cite{spiegel1972} reported a survey data of $n=320$ families with five children. The numbers of families with $0,1,2,3,4$ and $5$ girls were $18,56, 110, 88, 40$ and $8$. As an obvious model for $p_0$ we choose Binomial$(5, \hat \pi=.463)$. Estimated LP-Fourier coefficients are 
\beq  n \times \LP(p_0 \| \tp) \,=\,n \times \sum_{j=1}^5 \left|\LP[j;p_0,\tp]\right|^2 = 1.489~~\eeq
with $p$-value $0.92$ under the chi-square null with df $5$. In other words, we have just shown that the comparison density is flat uniform $d_0(u)=1$, hence the binomial distribution is completely acceptable for this data; no further density sharpening is needed.

\begin{rem}
Note that in our analysis, the prime object of interest is the shape of the estimated $\whd_0(x)$ (because it addresses Tukey's concern about the practical utility of goodness-of-fit), not how big or small the  $p$-value is. But if a data analyst is habituated to using a threshold $p$-value as a basis for decision making (not a good practice), then we recommend `double parametric bootstrap' \citep{beran1988} to compute the  $p$-value---
admittedly a computationally demanding task. This adjusted $p$-value takes into account the fact that the null-parameters are not given (e.g., here the binomial proportion $\pi$); they are estimated from the data.
\end{rem}
\vspace{-.1em}
\end{example}

 \begin{example}[Rutherford-Geiger polonium data] \label{ex:polo}
\cite*{rutherford1910} presented experimental data on groups of alpha particles emitted by Polonium, a radioactive element, in 1/8 minute intervals.  On the whole, $n =
2608$ time intervals were considered, in which $k$ ($k=0,1,\ldots,14$) decays were observed. The following table summarizes the data.
\begin{table}[ht]
\vskip.8em
\caption{Observed number of collisions of alpha particles emitted from polonium.} \label{tab:RFord}
\centering
\renewcommand{\tabcolsep}{.27cm}
\renewcommand{\arraystretch}{1.33}
\begin{tabular}{|rrrrrrrrrrrrrrr|}
  \hline
 & 0 & 1 & 2 & 3 & 4 & 5 & 6 & 7 & 8 & 9 & 10 & 11 & 13 & 14\\
  \hline
&  57 & 203 & 383 & 525 & 532 & 408 & 273 & 139 &  45 &  27 &  10 &   4 &   1 &   1 \\
   \hline
\end{tabular}
\vspace{.6em}
\end{table}
In the 1910 article, Bateman showed that (appended as a note at the end of the original 1910 paper) the theoretical distribution of alpha particles observed in a small interval follows Poisson law, which we select as our model-0. The estimated $\DS(p_0,m)$ model is given by
\beq \label{eq:rf}
\hp(x)\,=\,e^{-\la_0} \dfrac{\la_0^x}{x!}  \,\Big [ 1 -0.03 T_2(x;F_0)   -0.04 T_3(x;F_0) \Big ], ~~\text{with}~\la_0=3.88\eeq
which is displayed in Fig. \ref{fig:polo} in the Appendix. The model \eqref{eq:rf} indicates that there is a `gap' between the theoretically predicted Poisson model and the experimental result---
second-order (under‐dispersed) and third-order (less skewed) corrections are needed. To quantify the lack-of-fit, compute:
\[ n \times \LP(p_0 \| \tp) \,=\,n \times \sum_{j\in \{2,3\}} \left|\LP[j;p_0,\tp]\right|^2 = 6.82,~~\text{with pvalue $0.033$}. \]
This is a borderline case, where it is important to consult subject matter specialists before choosing sides---scientific significance is as important as statistical significance. \cite{hoaglin1980} came to a similar conclusion using an exploratory diagnostic tool called ``Poissonness plot,'' shown in the Appendix Fig. \ref{fig:poloeda}.  
\end{example}

\begin{example}[Sparrow data]
This data composed of numbers of sparrow nests found in plots of area one hectare, the sample average being $\bar x=1.10$. \cite{zarbook} previously analyzed this dataset. We choose  Poisson($1.10$) as our $p_0(x)$ for $\DS(p_0,m)$ model. The second-order $\LP[2;p_0,\tp]=-0.328$ (pvalue=.03) turns out to be the only  significant component, which indicates that the data exhibit less-dispersion (due to the negative sign) than postulated ${\rm Poisson}(1.10)$. Our finding is in agreement with \cite{gurtler2000}. Finally, return the $d$-modified under-dispersed Poisson model for the data:
\[\hp(x)\,=\,e^{-\la_0} \dfrac{\la_0^x}{x!}  \,\big [ 1 -0.33 T_2(x;F_0)\big ], ~~\text{with}~\la_0=1.10\] 
which is displayed in Fig. \ref{fig:sparrow} of the Appendix. 
\end{example}

\subsection{Relative Entropy and Model Uncertainty} \label{sec:rent}
How to quantify the uncertainty of the chosen model $p_0(x)$? A  general information-theoretic formula of model uncertainty is derived based on relative entropy between the true (unknown) $p(x)$ and the hypothesized $p_0(x)$. Express relative entropy (also called Kullback–Leibler divergence) as a functional of maxent comparison density $d_{\teb}$:
\beq
\KLD(p\|p_0)\,=\,\sum_x p(x) \log \Big\{\dfrac{p(x)}{p_0(x)}\Big\}\,=\,\Ex_F\Big[ \log d_{\teb}\big(F_0(X); p_0, p\big)\Big]
\eeq
Substituting $F_0(x)=u$, leads to the following important formula for relative entropy in terms of LP-parameters:
\bea 
\label{eq:LPrent}
\KLD(p\|p_0)&=&\int_0^1 d(u;F_0,F) \log d(u;F_0,F) \dd u \nonumber \\
&=&  \int_0^1 d(u;F_0,F)  \Big\{  \sum_j \te_j S_j(u;F_0)\,-\,  \Psi(\teb) \Big\}             \nonumber \\
&=&\sum_j \te_j \LP_j\,-\,\Psi(\teb). \label{eq:klp}
\eea
The second equality follows from  eq. \eqref{eq:lpeq} and the last one from eq. \eqref{eq:lpeq}. Based on a random sample $X_1,\ldots,X_n$, a nonparametric estimate of  relative entropy is obtained by replacing the unknown LP-parameters in \eqref{eq:klp} with their sample estimates.



\vskip.34em

{\bf Statistical Inference}. Relative entropy-based statistical inference procedure is applied to few real datasets in the context of model validation and uncertainty quantification. 
\vskip.1em
$\bullet$ Estimation and Standard error: For the earthquake data, we like to quantify the uncertainty of $p_0={\rm NB}(12, 19)$. The estimated value of $\KLD(p\|p_0)$ is $0.070 \pm 0.020$ (bootstrap standard error, based on $B=1000$), indicating serious lack-of-fit of the starting NB model---which matches with our previous conclusion; see Fig. \ref{fig:earthq}.
\vskip.2em
$\bullet$ Testing: For the Spiegel family data, the estimated relative entropy we get is $0.0087$, quite small. Naturally, we perform (parametric bootstrap-based) testing to check if $H_0: \KLD(p\|p_0)=0$. Generate $n$ samples from $p_0(x)$; compute $\widehat{\KLD}(p\|p_0)$; repeat, say, $1000$ times; return the $p$-value based on the bootstrap null-distribution. For this example, the $p$-value we get is $0.093$, which reaffirms that binomial distribution explains the data fairly well.

\subsection{Card Shuffling Problem}
\label{sec:card}
Consider the following question \citep{aldous1986}: How many times must a deck of cards be shuffled until it is close to random? To check whether a deck of $52$ cards is uniformly shuffled we use fixed point statistic, which is defined as the number of cards in the same position after a random permutation. Large values of fixed points (`too many cards left untouched') is an indication that the deck is not well-mixed. 

\vskip.23em
\textit{Theoretically-expected distribution}: One of the classical theorems in this direction is due to Pierre \cite{de1713essay} who showed that the distribution of the number of fixed-points under $H_0$ (random permutation of $\{1,2,\ldots,52\})$ is approximately $p_0={\rm Poisson}(1)$.

\vskip.23em

\textit{Data and notation}: Let $n$ denotes sample size and $k$ number of shuffles. Then \texttt{CARD}$(k,n)$ stands for a dataset $X_1,X_2,\ldots, X_n$, where $X_i$ is the number of fixed points of a $k$-shuffled deck. By $\widetilde p_k$, we mean the sample distribution of fixed-point statistic after $k$ random permutations. The goal is to find the minimum value of $k$, such that it is safe to accept $H_0: p_k=p_0$, where, we should recall, the null-distribution $p_0$ is ${\rm Poisson}(1)$.


\vskip.23em
\textit{Modeling via goodness-of-fit}: Fig. \ref{fig:card150} shows a \texttt{CARD}$(k,n)$ dataset with $k=150$ and $n=500$. There is a clear discrepancy between the observed probabilities $\tp_k$ and the theoretical $p_0$. The estimated $d$-sharpen Poisson$(1)$ is given below:  
\beq \label{eq:card150} \widehat p_k(x)= \dfrac{e^{-1}}{x\,!} \big[ 1+ 0.130 \,T_1(x;F_0)\big], ~~~~x=0,1,\ldots\eeq
which shows that `first order perturbation' (location correction) is needed: $\LP[1;p_0,\tp_k]=0.130$ with pvalue $0.003$. The positive sign of $\LP_1$ indicates that the mean of the fixed points distribution with $k=150$ is larger than the postulated $\la_0=1$; more  shuffling is needed to make the deck close to random. 
The shape of \eqref{eq:card150} is shown in Fig. \ref{fig:card150}.

\vskip.3em
\textit{New updated mean}. A curious reader might want to know precisely how large the mean of $p_k$ is compared to $1$. For a distribution $F \sim {\rm DS}(p_0,m)$, we can write an expression for the mean of $F$ ($\la_F$) in terms of the mean of $F_0$($\la_0$). In this case, we have 
\bea 
\vspace{-1em}
\la_{k} ~=~\Ex_{F_k}[X]&=& \int x \,\big[ 1+ 0.130 \,T_1(x;F_0) \big] \dd F_0(x) \nonumber\\
&=&\int_0^1 Q_0(u) \big[ 1+ 0.130 \,S_1(u;F_0) \big] \dd u \nonumber\\
&=&\int_0^1 Q_0(u) \dd u + 0.130 \,\langle Q_0, S_1 \rangle_{\cL^2[0,1]} \nonumber\\
&=& 1 + 0.130 \times 0.9596 \approx 1.125.
\eea

\vspace{-1.5em}
A few additional comments:

\begin{figure}[ ]
  \centering
  \vspace{-.8em}
\includegraphics[width=.55\linewidth,keepaspectratio,trim=2cm 1.65cm 2cm 1cm]{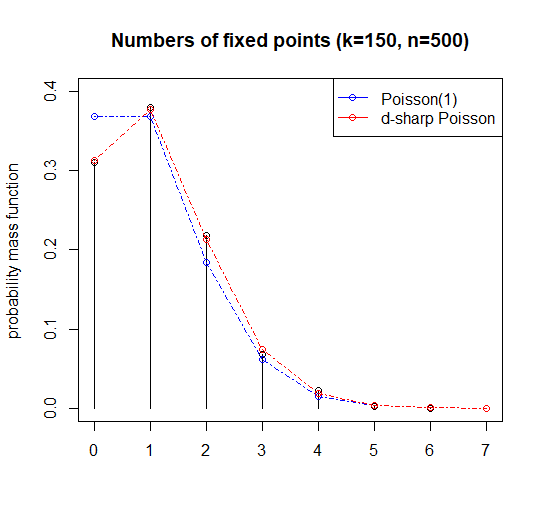}

\caption{Distribution of fixed points: black is empirical and blue is theoretical---a clear mismatch between data and the theory. The rectified $\DS(p_0,m)$ model is displayed in red.}\label{fig:card150}
\end{figure}

$\bullet$ Thus far, we have verified that $k=150$ is not enough to produce a  uniformly shuffled deck. However, we came to this conclusion based on a \textit{single} sample of size $n$. So, we generate (through computer simulation) several replications of \texttt{CARD}$(k,n)$ data with different $(k,n)$.

\vspace{-.15em}
$\bullet$ To reach a confident decision, we perform the experiment with $n=500$ and $k=150, 160,170, 180, 190, 200$. The analysis was done based on $B=250$ datasets from each $(n,k)$-pair. The results are summarized in appendix Fig. \ref{fig:card2}, which shows that $k=170$ shuffles is probably a safe bet to declare a deck to be fair---i.e., uniformly distributed.

\vspace{-.15em}
$\bullet$ 
\cite{diaconis2018bayGOF} describes a Bayesian approach to this problem. In contrast, we offered a completely nonparametric solution that \textit{leverages} the additional knowledge of the expected $p_0(x)$ and provides more insights into the nature of discrepancies.




\subsection{Jaynes Dice Problem}

The celebrated Jaynes' dice problem \citep{jaynes62} is as follows: Suppose a die has been tossed $N$ times (\textit{unknown}) and we are told only that the average number of the faces was $4.5$---not $3.5$, as we might expect from a fair die. Given this information (and nothing else), the goal is to determine probability assignment, i.e., what is the probability that the next throw will result in face $k$, for $k=1, \ldots,6$.

\vskip.25em
\textit{Solution of Jaynes' dice problem using density sharpening principle}. The initial $p_0(x)$ is selected as the discrete uniform distribution $p_0(x)=1/6$ for $x=1,\ldots, 6$, which reflects the null hypothesis of `fair' dice. As we are given only the first-order location information (mean is $4.5$) we consider the following $\DS(p_0,m=1)$ model:
\beq 
\label{eq:Jds}
p(x) = \dfrac{1}{6}\Big\{ 1 + \LP[1;p_0,p]\,T_1(x;p_0) \Big\} \eeq
The coefficient $\LP[1;p_0,\tp]$ has to be estimated, and for that we also need to know the basis function $T_1(x;F_0)$.

Step 1. To find an explicit formula for the discrete basis $T_1(x;F_0)$, apply \eqref{eq:T1} with $F_0(x)=x/6$ and $p_0(x)=1/6$
\beq 
T_1(x;F_0) \,=\, \sqrt{12} \dfrac{(x/6 - .5)}{\sqrt{1-\sum_{x=1}^6 (1/6^3)}}\,=\,\sqrt{\dfrac{12}{35}} \big(   x - 3.5\big),~~\text{for}\,x=1,\ldots,6.
\eeq 

Step 2. Compute  $\LP[1;p_0,\tp]$ by applying formula \eqref{eq:eestlp}
\beq 
\label{eq:lpJ}
\LP[1;p_0,\tp] \,=\, \sum_x \tp(x) T_1(x;p_0) \,= \,\sqrt{\dfrac{12}{35}} \big(  \sum_x x \tp(x) - 3.5\big) \,=\, 0.586.
\eeq
The non-zero $\LP[1;p_0,\tp]$ indicates it was a loaded die.

\begin{figure}[ ]
  \centering
\includegraphics[width=.55\linewidth,keepaspectratio,trim=2cm 1.65cm 2cm 1cm]{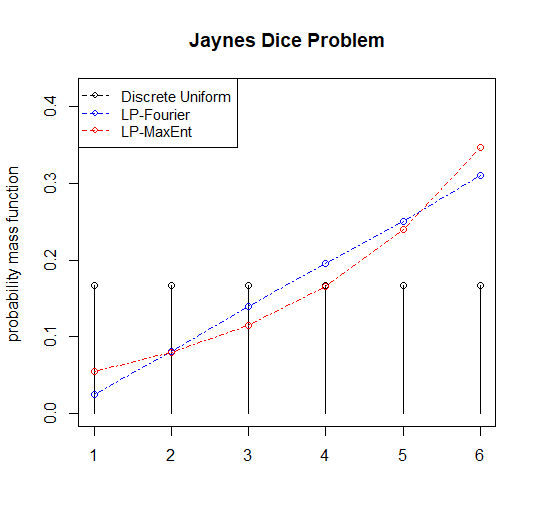}
\vskip.4em
\caption{The black dots denote the starting null-model---die with $6$ equally likely outcomes. The blue curve is our LP-Fourier density estimate given in eq. \eqref{eq:Jdses}. 
The exponential $\DS(p_0,m)$ density estimate, given in eq. \eqref{eq:maxentlp}, is shown in red line.}\label{fig:Jaynes}
\end{figure}

Step 3. Substitute the value of $\LP[1;p_0,\tp]$ in eq. \eqref{eq:Jds} to get the LP-Fourier $DS(p_0,m=1)$ model as 
\beq 
\label{eq:Jdses}
\widehat{p}(x) = \dfrac{1}{6}\Big\{ 1 + 0.586\,T_1(x;p_0) \Big\}, ~~x=1,\ldots,6. \eeq
This is shown as the blue curve in Fig. \ref{fig:Jaynes}.

Step 4. Finally, return the estimated exponential $\DS(p_0,m=1)$ probability estimates
\beq
\label{eq:maxentlp}
\hhp(x) (x)\,=\,\dfrac{1}{6}\exp\big\{ -0.193 + 0.634\,T_1(x;F_0) \big\},~~x=1,\ldots,6.
\eeq
This is shown as the red curve in Fig. \ref{fig:Jaynes}. 

\begin{table}[h]
\centering
\vskip1em
\caption{The estimated LP skew distribution for the Jaynes' dice problem.}
\vskip.5em
\renewcommand{\tabcolsep}{.3cm}
\renewcommand{\arraystretch}{1.77}
\begin{tabular}{c|cccccc}
 \hline
$p_0(x)$ &1/6&1/6&1/6&1/6&1/6&1/6\\
LP-Fourier $\hp(x)$ &.025&.080 &.140 &.195 &.250 &.310\\
LP-MaxEnt $\hhp(x)$ &0.054 &0.079 &0.114 &0.165&0.240&0.347 \\
Jaynes'Answer  &0.054 &0.079 &0.114 &0.165&0.240&0.347\\
\hline
\end{tabular}
\label{tab:Jaynes}
\vskip.3em
\end{table}

\begin{rem}
It is a quite remarkable fact that our density-sharpening principle-based probability assignment \textit{exactly} matches with Jaynes' maxent answer; see Table \ref{tab:Jaynes}.
\vspace{-.25em}
\end{rem}




\subsection{Compressive Learning of Big Distributions}
An important class of learning problem that has recently attracted researchers from various disciplines---including high-energy physics,  neuroscience, theoretical computer science, and machine learning---can be viewed as a modeling problem based on samples from a distribution over large ordered domains. Let ${\bf p}=(p_1,\ldots,p_k)$ be a probability distribution over a very large domain $k$, where $p_i \ge 0, \sum_{i=1}^k p_i =1$.  Let us look at a realistic example before discussing a general method.

\begin{figure}[ ]
\vspace{-.6em}
  \centering
\includegraphics[width=.478\linewidth,keepaspectratio,trim=1cm 1cm .1cm 1.2cm]{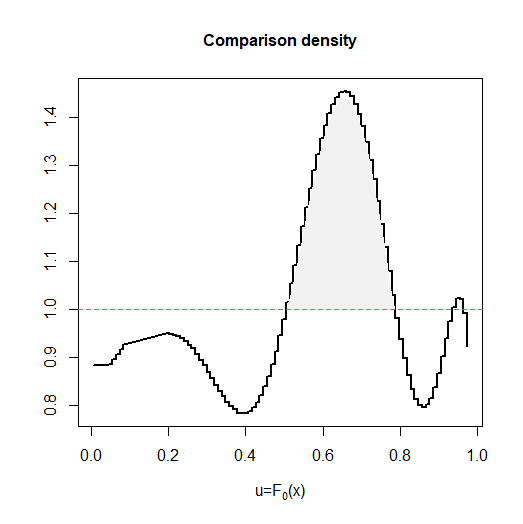}~~~~~
\includegraphics[width=.48\linewidth,keepaspectratio,trim=.35cm 1cm 1cm 1cm]{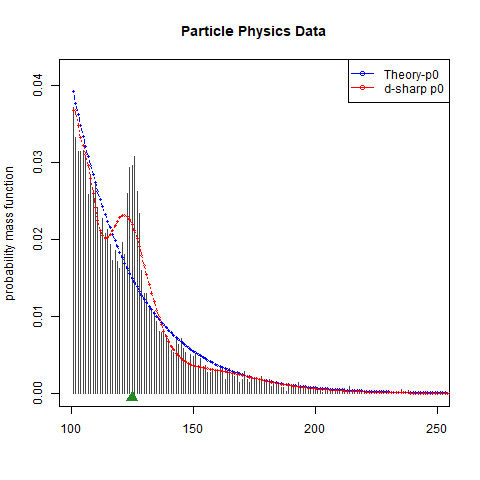}
\vskip.6em
\caption{\texttt{HEP}$(k,n)$ data analysis. Left: The estimated comparison density $\whd(u;p_0,p)$, which indicates there may be a bump (around $u=0.64$) that went \textit{unnoticed} by the theoretical model. Right: The data generated from a mixture of $p_0$ and $\cN(125, 2)$ with mixing proportion $0.1$. The theory-driven model $p_0$ is the blue curve and the red curve is the $d$-sharpen $p_0$. The green triangle denotes the true peak at the mass position 125 Gev. The shaded area under $\dhat_0(u)$ denotes the amount of excess mass on the top of the smooth background.}\label{fig:PP}
\end{figure}

\vskip.45em
\begin{example}[HEP data]
This is an example from high-energy physics\footnote{High-energy physics is not the only discipline where this kind of very large and sparse histogram-like data appears. It frequently arises in many modern scientific domains: inter-spike interval data (neuronal firing patterns); relative abundance/intensity data (mass spectrometry data); DNA methylation and ChIP-seq data  (genomics); pixel histogram data (astronomical images of stars, galaxies etc); histograms of activity intensities (biosignals from wearable sensor devices, mental illnesses studies by NIMH); photometric redshift data (photo-z spectra in Cosmology),  just to name a few. There is an outstanding interest in developing new computational tool that allows rapid and approximate statistical learning for big-histogram-like datasets.} (HEP), motivated by the PHYSTAT 2011 Banff bump-hunting challenge task \citep{junk2011banff}. In HEP counting experiments (e.g., in Large Hadron Collider) one observes data in the following form: $n$ samples from unknown $p(x)$ as event counts (number of collisions) at $k$ finely binned energy-cells, which we denote by \texttt{HEP}$(k,n)$. Fig. \ref{fig:PP} displays one such data with $n=10,000$ and $k=250$, with the postulated background model (dictated by the known Standard Model) as discretized exponential distribution $f_0(x)=\lambda e^{-\lambda x}$ with $\lambda=1/20$:
\beq 
\label{eq:nullHEP}
p_0(i) \, \doteq \, \int_{x_i}^{x_{i+1}} f_0(x) \dd x,~~i=1,\ldots,k.\eeq
Particle physicists are interested in discovering new particles that go \textit{beyond} the known Standard model described by the background model $p_0$.  We present a four-step algorithmic program to address the general problem of data-driven `Learning and Discovery.'
\end{example}




\vskip.3em
{\bf Phase 1.} \textit{Testing}. The first question a scientist would like answered is whether the data is consistent with the background-only hypothesis, i.e., $H_0:p=p_0$.  We perform the information-theoretic test described in Sec. \ref{sec:rent}. In particular, we choose the relative entropy-based formula given in \eqref{eq:LPrent} as our test statistic. The  $p$-value obtained using parametric bootstrap (with $B=50,000$) is almost zero---strongly suggesting that the data contain some surprising new information in light of the known physics model $p_0$. But to figure out whether that information is actually useful for physicists, we have to dig deeper.

\vskip.15em

{\bf Phase 2.} \textit{Exploration and Discovery}. By definition, new discoveries can be made only by ``contrasting'' data with the existing model. This is what is achieved through $d(u;F_0,\wtF)$. The left panel of Fig. \ref{fig:PP} displays the estimated $\whd_0(u)$ for the HEP-data, which compactly encodes all the structure of the data that cannot be described by the assumed $p_0(x)$. 

\vskip.15em
The exploratory graphical display of $\whd_0$ reveals a few noteworthy points. Firstly, the non-uniformity of $\whd_0$ makes us skeptical about $p_0$---this is completely in agreement with Phase-1 analysis.  Secondly and more importantly, the shape of $\whd_0$ provides a refined understanding of the \textit{nature} of new physics that is hidden in the data, which, in this case, revealed itself as a bump \textit{above} the smooth background $p_0$. The word `above' is important because we are not interested in bumps on $p$ itself, but on $d_0$, which is the unanticipated `excess mass.' Hunt for new physics is the problem of bump hunting on $d(u;p_0,p)$, not on $p(x)$. For the HEP-data, we see a prominent bump in $d_0(u)$ around $u=0.64$, which (in the original data domain) corresponds to near $Q_0(.64) \approx$ 125 GeV; the green triangle in the above figure. 

\begin{rem}[The discovery function]
Since, $d_0(x)$ encapsulates what's new in the data by separating the unknown from the known, we also call it the ``discovery function.'' It is the ``missing piece'' that glues together the known $p_0(x)$ and the unknown $p(x)$. 
It provides a graphical diagnostic tool that exposes the unexpected. These clues can guide domain-scientists to carry out more targeted follow-up studies. 
\end{rem}


\vskip.3em
{\bf Phase 3.} \textit{Inference and Excess Mass Problem}.  Where is the interesting excess mass hiding? Is it a statistical fluke or something real? How substantial is the evidence?   The real issue is: can we let the data confidently tell us where to look next for new particles? This will result in a complete paradigm shift because traditionally the HEP searches (for locating excess events) were guided by theoretical considerations only.
\begin{quote}
\vspace{-.3em}
    `\textit{One may feel uneasy that we may therefore only find new processes if a theorist has been clever enough to propose the corresponding theory ahead of time.}'
\begin{flushright}
\vspace{-.24em}
{\rm ----Glen \cite{cowan2007}} \end{flushright}
\end{quote}
\vspace{-.3em}
Interested readers may also refer to \cite{lyons08} and the Nature news article by \cite{castelvecchi2018lhc} for a clear exposition on the scientific importance of these issues.

\begin{figure}[ ]
  \centering
\includegraphics[width=.48\linewidth,keepaspectratio,trim=2.5cm 1cm 2.5cm 1cm]{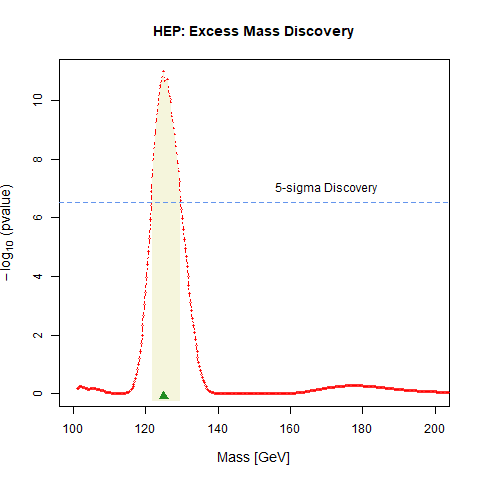}
\vskip2em
\caption{ The $5\sigma$ discovery interval $(121.5,129.5)$ correctly captures the true excess-mass point indicated by the green triangle.}\label{fig:HEPxm}
\end{figure}

\vspace{.24em}
{\bf Statistical Discovery: Inference Algorithm}. The following are the main steps of the inference algorithm whose results are summarized in Fig. \ref{fig:HEPxm}:

Step 1. Parametric bootstrap: To measure the natural statistical variation of $\dhat_0(x)$ under the null hypothesis: simulate $n$ samples from $p_0(x)$ and estimate the comparison density. Repeat the whole process for a large number of iterations (say $B=10,000$ times) to get a bundle of comparison density curves, all of which fluctuate around the flat uniform line.

Step 2. Pointwise $p$-value computation: At a fixed grid point $x \in [100, 250]$, we have the following values of the test statistic
\[\Big\{ \whd_0^{(1)}(x), \ldots,  \whd_0^{(B)}(x)  \Big\} \]
calculated from the $B$ bootstrap samples. Compute the bootstrap $p$-value at the point $x$ by
\[{\rm Pval}(x)\,=\,\dfrac{1+ \big\{ \# \,{\rm of}\,  \whd_0^{(j)}(x) \ge  \whd_0(x)\big\}   }{B+1 }\]
Fig. \ref{fig:HEPxm} draws the curve $-\log_{10}({\rm Pval}(x))$ as a function of $x$. 
The $5\sigma$ discovery region $(121.5, 129.5)$ is highlighted in yellow, which includes the true excess mass point $125$ GeV. This is how modern nonparametric modeling based on `density sharpening principle' can convincingly guide researchers on \textit{where} to look for evidence of a deeper theory of physics.

\vskip.3em
{\bf Phase 4.} \textit{Sharpen Scientific-Model}. Finally, the goal is to sharpen the initial scientific model $p_0(x)$ to achieve a more precise description of what is loosely known or suspected. The estimated $\DS(p_0,m)$ model sharpens the parametric null \eqref{eq:nullHEP} to provide a nonparametrically-adjusted, parsimonious model:
\beq 
\label{eq:dshep}
\hp(x)\,=\,p_0(x)\Big[  1 -\sum_{j \mathcal{J}_5} \LP[j;p_o,\tp]\,T_j(x;F_0)\Big],
\eeq
where the active set $\mathcal{J}_5=\{2,3,5,7,8\}$ along with the LP-coefficients are given in Table \ref{tab:HEP}.
\begin{table}[ht]
\vskip.4em
\caption{The selected no-zero LP-coefficients}
\label{tab:HEP}
\centering
\renewcommand{\tabcolsep}{.4cm}
\renewcommand{\arraystretch}{1.33}
\begin{tabular}{|l |ccc cc|}
  \hline
$\mathcal{J}_5$    & 2& 3& 5& 7 &8\\
  \hline
 $\widehat{\LP}_j$   & -.097 & -.090& 0.117& -.095 & -.060 \\
   \hline
\end{tabular}
\end{table}
\vspace{-.3em}
\begin{rem}
A few remarks:

~1. The part in the square brackets of \eqref{eq:dshep} shows \textit{how} to change the prior scientific model $p_0(x)$ to make it consistent with the data. Knowing the \textit{nature} of the deficiency of the assumed model, is an important step towards data-driven discovery. As George \cite{box2001dis} said: ``\textit{discovery usually means learning how to change the model}.''

~2. LP-parameterization requires only $5$-dimensional sufficient statistics to approximately capture the shape of the distribution! The ``compressiveness'' of the LP-transformation---the ability to extract a low-dimensional representation---makes it less data-hungry, as demonstrated in the next section.

~3. Our model \eqref{eq:dshep} is a `hybrid' between theory-driven and data-driven model, which decouples the overall density into two components: expected $p_0(x)$ and unexpected $d_0(x)$. 
\end{rem}

\begin{rem}[An Appeal to Physicists: Hypothesis Testing $\neq$ Discovery Science]
Classical statistical inference puts too much emphasis on testing, $p$-value, standard error, and confidence intervals, etc. This ideology is reflected in the practice of high-energy physicists---which entirely revolves around antique tools of hypothesis testing, likelihood ratio, and $p$-value. It's time to break the shackles of outdated data analysis technology that starts with hypothesis testing and ends with a  $p$-value. George \cite{box2001dis} expressed a similar sentiment, arguing that the reason why engineering and the physical sciences rarely use statistics is: ``\textit{Much of what we have been doing is adequate for testing but not adequate for discovery.}''

\vskip.13em

In this section my purpose has been to introduce some modern statistical tools and concepts that can help scientists with their everyday tasks of discovery and deeper exploration of data. After all, one of the main goals of data analysis is to sharpen the scientists' mental model by revealing the unexpected---a continuous cycle of knowledge refinement:
\vskip2.5em
\begin{center}
\begin{figure}
\begin{tikzpicture}[node distance =4cm, auto]
    \node [block] (x1) {Theory};
    \node [block, right of    =x1] (x2) {Measurement};
    \node [block, right of    =x2](x3){Discovery $d_0(x)$};
    \node [block, right of    =x3](x4){Better theory};
   
    \path [line] ($(x1.0)+(.1cm,0cm)$)--($(x2.180)+(-.1cm,0cm)$);
    \path [line] ($(x2.0)+(.1cm,0cm)$)--($(x3.180)+(-.1cm,0cm)$);
    \path [line] ($(x3.0)+(.1cm,0cm)$)--($(x4.180)+(-.1cm,0cm)$);
    \path[line] ($(x4.south)+(0,-.1)$) -- +(0, -2em) -| ($(x2.south)+(0,-.1)$);
\end{tikzpicture}
\vskip1.5em
\caption{Continuous learning by iterative model refinement: It develops increasingly `better' theory that explains new phenomena by broadening the scope of the previous theory.}
\end{figure}
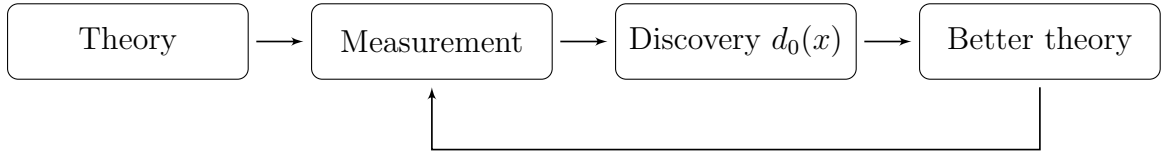
\vspace{-2em}
\end{center}





\end{rem}
\vspace{-.5em}

\subsection{Data-efficient Statistical Learning}\label{sec:DEL}
We are interested in the following statistical learning problem:  Given a small handful of samples $n \ll k$ from a big probability distribution of size $k$, how to perform time-and-storage-efficient learning? When designing such algorithms we have to keep in mind that they must be (i)  data-efficient: can learn from limited sample data $n \ll k$; and (ii) statistically powerful: can detect ``small'' deviations.

\vskip.3em
Classical nonparametric methods characterize big probability distributions using high-dimensional sufficient statistics based on histogram counts $\{N_j\}_{j=1}^k$, which, obviously, requires a very large sample for efficient learning. And as the required sample sizes increases, this slows down the algorithm running-time which scales with $n$.
Hence, most of the `legacy' statistical algorithms (e.g., Pearson's chi-square, Freeman-Tukey statistic, etc.) become unusable on large-scale discrete data as `the minimum number of data points required to obtain an acceptable answer is too large to be practical.'  Indeed, detecting sparse structure in a data-efficient manner from big distributions, such as from \texttt{HEP}($k,n$), is a challenging problem and requires a ``new breed'' of computational algorithms. There has been some impressive progress on this front by the Theoretical Computer Science community; see Appendix \ref{app:subl} for a related discussion on sub-linear algorithms for big data.
 
\begin{figure}[ ]
\vspace{-1em}
  \centering
\includegraphics[width=.5\linewidth,keepaspectratio,trim=2.25cm 1cm 2.25cm 1.25cm]{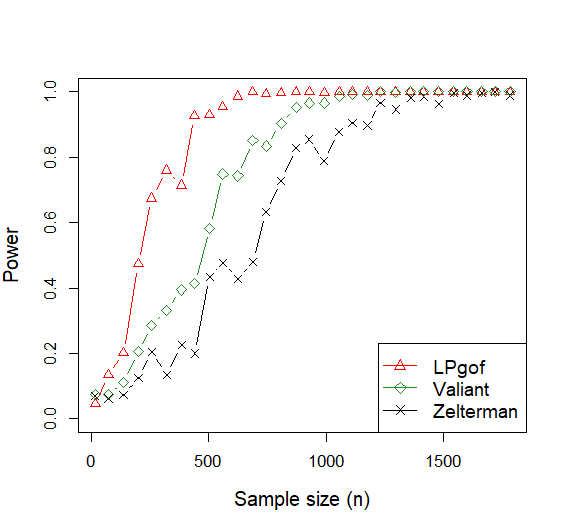}
\vskip.7em
\caption{HEP$(n,k=500)$ data for varying sample size $n$. Event counts (number of collisions) at $k = 500$ finely binned energy-cells. \texttt{LPgof} (shown in red) requires 50\% less data to reach the correct conclusion with power 1.}\label{fig:HEPower}
\end{figure}

\vskip.3em
{\bf HEP Data Example}.  We generate \texttt{HEP}($k=500,n$) data for varying sample size $n$. We used $350$ null simulated
data sets to estimate the 95\% rejection cutoffs for all methods at the significance level $0.05$,
and used $350$ simulated data sets from alternative to approximate the power, as displayed in Fig. \ref{fig:HEPower}. We compared our LPgof method \eqref{eq:ugof} with two state-of-the-art algorithms (proposed by theoretical computer scientists): (i) \cite{valiant2017automatic} and (ii) \cite{acharya2015optimal}---interestingly, this exact test has been proposed earlier by \cite{zelterman1987}, which in Statistics literature is known as Zelterman's D-statistic\footnote{`Those who ignore Statistics are condemned to reinvent it'---Brad Efron.}. Conclusion: \texttt{LPgof} requires {\bf 50\% less} data to reach the correct conclusion with power 1. 

\begin{figure}[ ]
\vspace{-.7em}
\centering
\includegraphics[height=.26\textheight,width=\textwidth,keepaspectratio,trim=1cm 1cm 1cm 1cm]{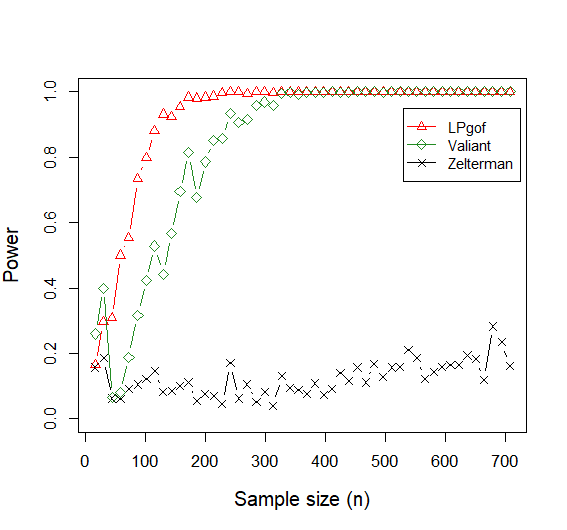}~~~~~~~~
\includegraphics[height=.26\textheight,width=\textwidth,keepaspectratio,trim=1cm 1cm 1cm 1cm]{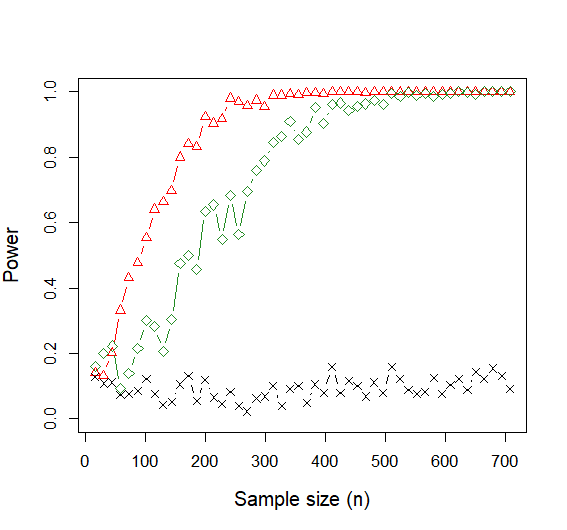} \\[1em]
\includegraphics[height=.26\textheight,width=\textwidth,keepaspectratio,trim=1cm 1cm 1cm 1cm]{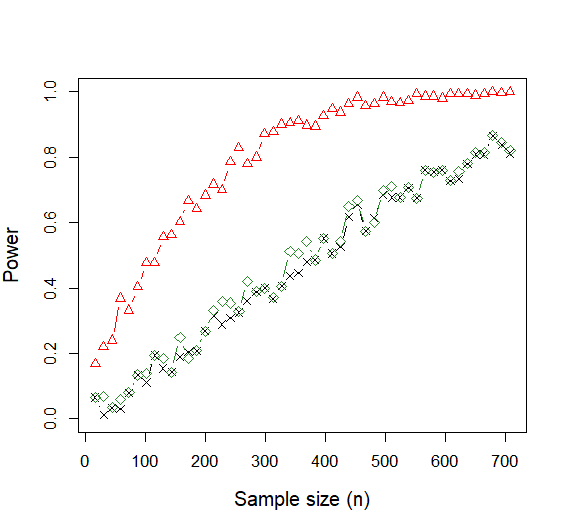}~~~~~~~~
\includegraphics[height=.26\textheight,width=\textwidth,keepaspectratio,trim=1cm 1cm 1cm 1cm]{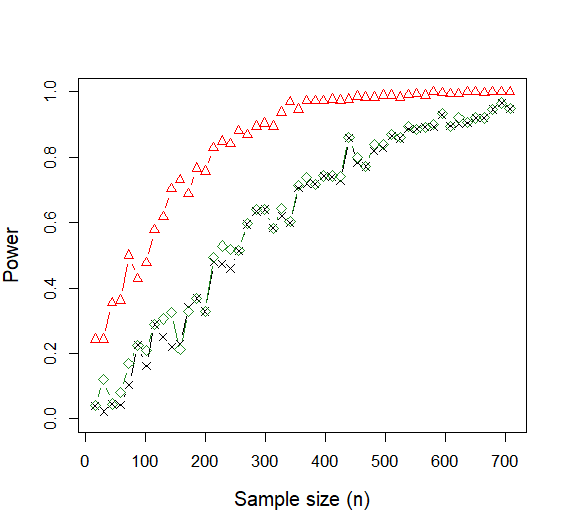} \\[1em]
\includegraphics[height=.26\textheight,width=\textwidth,keepaspectratio,trim=1cm 1cm 1cm 1cm]{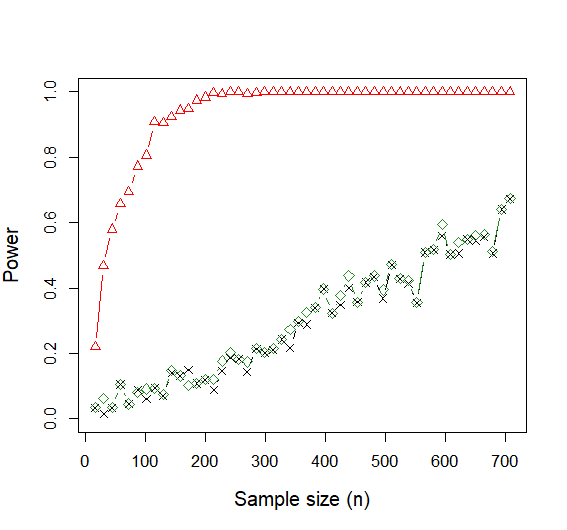}~~~~~~~~
\includegraphics[height=.26\textheight,width=\textwidth,keepaspectratio,trim=1cm 1cm 1cm 1cm]{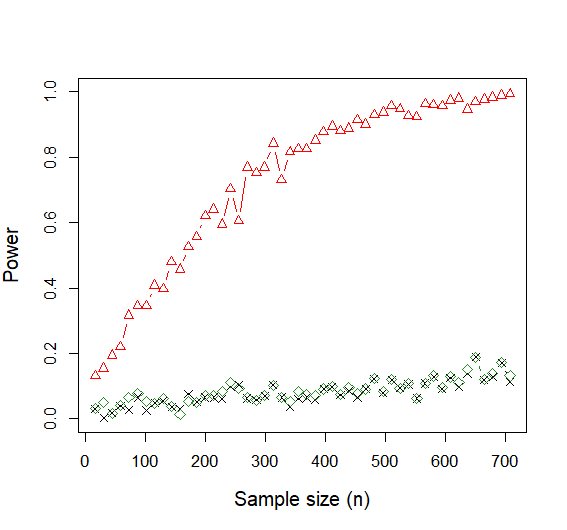} \\[1em]
\caption{The red power curve is \texttt{LPgof} method with $m=8$. First row:  Null distribution is $p_0=(1/4,1/4,1/2(k-2) \ldots, 1/2(k-2))$. Alternatives are $p_{\delta}=(1/4+\de,1/4-\de,1/2(k-2) \ldots, 1/2(k-2))$, with $\de=1/8,1/10$. Second row: Null distribution is $p_0=U_{[k]}$. Alternatives are $.95U_{[k]} + .05{\rm Zipf}(\al)$, with $\al=1.25,1.15$. Third row: Null distribution is $p_0=U_{[k]}$. Alternative probabilities are computed using the increments $\Delta D(j/k; \mu,\pi)$, $j=1,\ldots,k$ where $D(u)=F(\Phi^{-1}(u);X)$, and $F(x;X)=(1-\pi)\Phi(x) + \pi \Phi(x-\mu)$ for $\mu=-1.5$ and $\pi=\{.2,.1\}$. In all cases the dimension $k=5000$; and sample sizes are sublinear in dimension, i.e., $n \asymp \sqrt{k}$} \label{fig:power1}
\end{figure}

{\bf Empirical Power Comparisons}. We compare the power of different methods under six different settings, as described in the Figs \ref{fig:HEPower} description, and will not repeat this here. The overall conclusion is pretty clear: \texttt{LPgof} emerged as the most powerful data-efficient test---it can detect new discoveries \textit{quickly and more reliably}. The prime reason for achieving this level of performance is fully attributable to the good sparsity (energy compaction) property of ``LP-domain data analysis." The specially-designed discrete LP-transformation basis provides an efficient coordinate system that requires far fewer parameters than the size of the distribution to capture the essential information.  For additional examples see Figs. \ref{fig:powerh0} and \ref{fig:power2}
of the Appendix. 


\subsection{Discovery-source Separation Problem}
~~``\textit{At this scale it is not possible to keep all the data (the LHC produces up to a Petabyte of data per second) and it is essential to have efficient data-filtering mechanisms so that we can separate the wheat from the chaff.}''
\begin{flushright}
\vspace{-1.35em}
{\rm --- Bob Jones, Project Leader at CERN.}\end{flushright}

\begin{figure}[ ]
  \centering
  \includegraphics[width=.46\linewidth,keepaspectratio,trim=2.25cm 1cm 1cm 1cm]{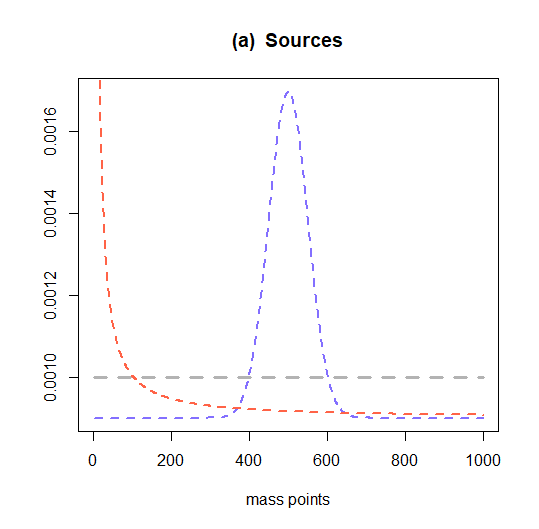}~~~~
\includegraphics[width=.46\linewidth,keepaspectratio,trim=1cm 1cm 2.25cm 1cm]{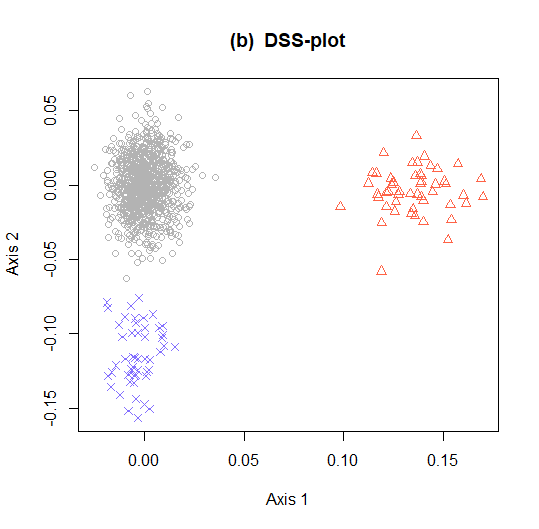}
\vskip1em
\caption{Discovery-source Separation (DSS) plot.
Each data-batch is HEP$(k=1000,n=1000)$ with different shapes, among them $900$ were generated from $U_{[k]}$ (grey dotted line), $50$ were from $.9U_{[k]} + .1{\rm Zipf}(1.15)$ (red dotted line), and $50$ from a mixture of $U_{[k]}$ and increments of ${\rm Beta}(50,50)$ with mixing proportion $.1$ (blue dotted line). In the right panel, we show the DSS-plot. The chunks of points around zero indicate the null data sources. The sources are classified into three groups based on their nature of discoverability.}\label{fig:dss}
\end{figure}

Imagine a typical scenario, where massive experimental data are distributed across thousands of different computation units (say, servers). Each data source has its own `personal' distribution over a large domain of size $k=1000$, from which, we only have access to $n=1000$ samples. Thus, the observed data can be summarized as a collection of highly noisy and sparse empirical distributions $\{\tp_\ell(x)\}$. Suppose we are given $900$ such sparsely-sampled empirical distributions from $p_0=U_{[k]}$, $50$ from $.9U_{[k]} + .1 {\rm Zipf}(1.25)$, and $50$ from 
a mixture of $.9U_{[k]} + .1 \Delta{\rm Beta}(50, 50)$, where $\Delta{\rm Beta}(50, 50)$ denotes the increments of the cdf of ${\rm Beta}(50,50)$. Shapes of these three source-distributions are shown in the left of Fig. \ref{fig:dss}.

\vskip.35em

{\bf Discovery-source Separation (DSS)}. The goal here is to design algorithms that can quickly filter out the `interesting' data sources, having potential for a new physics discovery. A more ambitious goal would be to classify different data sources based on their `nature' of discoverability.

\vskip.35em
{\bf Algorithm}. It consists of two main steps:
\vskip.2em
Step 1. LP-Fourier Transform Matrix: Define the LP-transform matrix $L \in \cR^{g\times m}$ by
\beq 
L[\ell,j]=\LP[j;p_0,\tp_\ell] = \int T_j(x;F_0) \dd \wtF_\ell,
\eeq
for $\ell=1,\ldots,g=1000$ and $j=1,\ldots, m=10$.
\vskip.25em

Step 2. DSS-plot: Perform the singular value decomposition (SVD) of $L=U\Lambda U^{T}$ $= \sum_r \la_r u_r u_r^{T}$, where $u_{ij}$ are the elements of the singular vector matrix $U=(u_1,\ldots,u_m)$, and $\Lambda={\rm diag}(\la_1,\ldots,\la_m)$, $\la_1 \ge$ $ \cdots \la_m \ge 0$. DSS-plot is the two-dimensional graph in the right panel of Fig \ref{fig:dss} (b), which is formed using the points  $(\la_1 u_{1\ell}, \la_2 u_{2\ell})$, for $\ell=1,\ldots,g$ by taking the top two dominant singular vectors. Different data sources are shown as points, which captures the heterogeneity in terms of discoverability.


\vskip.4em
\textit{Interpretation}: DSS-plot displays and compares a large number of (empirical) distributions $\tp_\ell$ by embedding them in 2D Euclidean space. The cluster of points (data sources) that are near to the origin are the ones with background distribution. The distance from the origin
\[\text{discovery-index}_\ell=(\la_1 u_{1\ell} -0)^2\,+\, (\la_2 u_{2\ell} -0)^2,~~~~\ell=1,\ldots,g=1000~~\]
can be interpreted as the ``degree of newness'' of that dataset. The DSS-plot successfully separates different sources based on the statistical nature of their signal components. Researchers (like Bob Jones) can use this tool to \emph{quickly identify} interesting data sets for careful investigation.




\section{Discussion} \label{sec:diss}
Compared to the rich and well-developed tools for continuous data modeling problems, many companion discrete data modeling problems are still quite challenging. This was the main motivation for undertaking this research, which aimed at developing a widely-applicable general theory of discrete data modeling. This paper makes three broad contributions to the field of  nonparametric statistical modeling:


\vskip.5em

~~1) Model-Sharpening Principle: We have introduced a new principle of statistical model building, called `Density-Sharpening,' which performs three tasks in one step: model verification, model exploration, and model rectification. This was the guiding principle behind our systematic theory of discrete data modeling. As future work, we plan to explore how model-sharpening principle can help developing `auto-adaptable' machine-learning models.

\vskip.5em

~~2)  $\DS(p_0,m)$ Model: We have introduced a new class of nonparametric discrete probability model and robust estimation techniques that has the capability to leverage researchers' vague (misspecified) prior knowledge. It comes with novel exploratory graphical methods for `discovering' new knowledge from the data that investigators neither knew nor expected.

\vskip.5em
~~3) Unified Statistical Framework:   Our modern nonparametric treatment of the analysis of discrete data is shown to be rich enough to subsume a large class of statistical learning methods as a special case.  This inclusivity of the general theory has some serious implications: Firstly,  from a theoretical angle, it deepens our understanding of the ties between different statistical methods. Secondly, it simplifies practice by developing unified algorithms with expanded capabilities. And finally, it is also expected to be beneficial for modernizing Statistics curriculum in a way that is applicable to small- as well as large-scale problems.

\section*{Supplementary Appendix}
\label{SM}
The supplementary material includes additional theoretical and algorithmic details.













\bibliographystyle{Chicago}
\bibliography{ref-bib}

\newpage

\section{Appendix}
\renewcommand{\theequation}{6.\arabic{equation}}
\appendix
\renewcommand{\thesubsection}{A.\arabic{subsection}}
\vskip.6em
It consists of four sections.

\subsection{Goodness-of-fit Methods: Automatic verses Manual} \label{app:gof}

Traditional GOF methods, mainly inspired by \cite{lan1953recona}, construct orthonormal polynomials on a \emph{case-by-case basis} for each parametric discrete distribution $p_0(x)$. This is generally done by solving the heavy-duty Emerson recurrence \citep{emerson1968a,best1999a,best2003a} relation, which could be quite complicated for non-standard distributions. Few concrete examples: The proposal of \cite{best1997bina} for testing Binomial distribution is based on Krawtchouk polynomial; \cite{best2003testsa} develop test for geometric  distribution using Meixner orthonormal polynomials; \cite{best1999poisa} construct Poisson GOF method based on  Poisson-Charlier orthonormal polynomials; For a theory of classical (distribution-specific) orthonormal polynomials refer to \citet{chihara2011booka}.

 
\vskip.32em
The problem with this line of thinking is that it makes the whole procedure hard to automate and generalize, which may turn off practitioners.


\vskip.32em
Instead of bookkeeping a long list of esoteric polynomials for each parametric distribution, we want to devise a universal mechanism that works for any arbitrary $p_0$---which will make it easy to use and compute. This is achieved in Section \ref{sec:Xgof} of our paper by employing LP-orthonormal polynomial bases. Its universality \eqref{eq:ugof} lies in the fact that, unlike traditional methods, we have the ability to construct discrete orthonormal polynomials (Sec. \ref{sec:LPFA}) of the user-specified $p_0(x)$ in a completely \textit{automatic and robust} manner. 






\vskip1em
 
\subsection{LPdiscrete: A Unified Algorithm} \label{app:algo}
We summarize some of the key steps of our generic discrete data modeling approach, which starts with data $\{X_1,\ldots,X_n\}$ and an approximate (misspecified) working model $p_0(x)$.
\newpage 

\begin{center}
\texttt{LPdiscrete}: Algorithm Based on `Density-Sharpening' Principle
\end{center}
\vspace{-.4em}
\medskip\hrule height .8pt
\vskip1em
1. Input: Discrete sample distribution $\tp(x)$ and the assumed reference model  $p_0(x)$. The corresponding cdfs are denoted by $\wtF(x)$ and $F_0(x)$.
\vskip.6em
2. Compute the first-order LP-basis $T_1(x;F_0)$ by standardizing mid-distribution transform
\[T_1(x;F_0) \,=\,\dfrac{\sqrt{12} \big[\Fmn(x) - 0.5\big]}{\sqrt{1-\sum_x p_0^3(x)}}.\]
\vskip.24em

3. Apply \emph{weighted} Gram-Schmidt procedure on $\{T_1^2, T_1^3,\ldots\}$ to construct empirical orthonormal polynomials $T_j(x;X)$ with respect to measure $F_0$
\[\sum\nolimits_x p_0(x) T_j(x;F_0)=0;~~\,\sum\nolimits_x p_0(x) T_j(x;F_0)T_k(x;F_0)=\delta_{jk}, ~~1<j,k<M \]
Define the Unit LP-bases $S_j(u;F_0)=T_j(Q_0(u);F_0)$, $0\le u \le 1$.
\vskip.6em

4. Estimate the discrete LP-Fourier transform coefficients of $p(x)$ with respect to $p_0(x)$
\[\tLP_j\,:= \LP[j;p_0;\tp]\,=\,  \Ex\big[T_j(X;F_0);\widetilde F\big]\,=\,\sum_x \tp(x) T_j(x;F_0), \]
They are the coordinates of the true unknown distribution $F$ relative to $F_0$ 
\[\big[ F \big]_{F_0} := \Big(\LP[1;F_0,F], \ldots, \LP[m;F_0,F]\Big).\]

5. Sparse LP-transform. Identify the `significant' non-zero LP-coefficients with $|\tLP_j|>2/\sqrt{n}$. A more  sophisticated method: identify indices $j$ using AIC (or BIC) model selection criterion applied to $\tLP_j$ arranged in decreasing magnitude
\[{\rm AIC}(m)~=~\textrm{Sum of squares of first m coefficients}\, – \, 2m/n.\]
Choose $m$ to maximize $\operatorname{AIC}(m)$. Store the selected indicates in the set $\cJ$.
\vskip.4em
6. Compute the lack-of-fit of the assumed model $p_0$
\[\LP(p_0 \| \tp)\,=\sum_{j\in \cJ} \Big|  \LP[j;p_0,\tp] \Big|^2= \sum_{j\in \cJ} \Big|\sum_x \tp(x) T_j(x;F_0)\Big|^2.\]
7. Since $\LP(j;p_0,\tp)$ is the inner product of comparison density $\dd \widetilde F/\dd F_0$ with LP-basis $T_j(x;F_0)$, they are the coefficient of the orthogonal expansion of $d_0(x)$ in LP-polynomial bases. Thus, the estimated LP-Fourier comparison density is given by:
\[\whd_0(u) := d\big(u;F_0,\widetilde F\big)\,-\,1\,=\,\sum_{j \in \cJ} \tLP_j\, S_j(u;F_0) .\]
8. Construct MaxEnt comparison density model based on selected LP-sufficient statistics 
\[ \whd_{\teb}(u;F_0,F)\,=\,\exp\Big \{ \sum_{j \in \cJ} \widehat{\te}_j S_j(u;F_0)\,-\, \Psi(\teb)\Big \},~~0<u<1\]
9. Estimate the relative entropy ${\rm KL}(p_0\|p)$ using the LP-parameter-based formula:
\[\sum_{j \in \cJ} \widehat{\te}_j \tLP_j\,-\,\Psi(\teb)\]
This quantifies the uncertainty that surrounds the initial tentative model $p_0(x)$.
\vskip.5em
10. Exploratory uncertainty analysis. The graphical visualization of $\whd(u;F_0,F)$ plays a central role in data-driven knowledge discovery. It conveys the \textit{missing} knowledge that was not anticipated by the a-priori assumed $p_0(x)$---tool for detecting the unexpected. It also provides insights as to \textit{how to change} the misspecified $p_0$ to get a better model that fits the data well. The purpose of this exploratory interface is to encourages interactive data analysis and visual reasoning.
\vskip.6em
11. Model amendment via density-sharpening: Construct a class of reasonable models in the neighborhood of the reference model $p_0(x)$
\[\hp(x)~=~p_0(x) \times \whd_0(x),\]
for $\whd_0(x)$ either use LP-Fourier canonical estimate or the LP-maxent estimate. 
The bottom line: Density-sharpening provides a systematic statistical model developmental process, describing 
how a relatively simple model $p_0(x)$ can \textit{evolved} into a more mature model $\widehat{p}(x)$ when it comes in contact with new data. The $\hp(x)$ is not a conventional nonparametric model, it is a `hybrid' model that \textit{supplements} domain-knowledge with data-knowledge.

\vskip1em
\medskip\hrule height .8pt
\vskip1em

\vskip1em
\subsection{Sub-linear Algorithms for Detecting Sparse Structure} \label{app:subl}
The problem of developing `efficient' algorithms for detecting sparse structure in big discrete distributions has recently attracted lots of attention from the Theoretical Computer Science (TCS) community.   By `efficient' we mean algorithms that require less data to correctly detect the presence of signals. We call an algorithm sub-linear if the sample complexity ($n$) is sub-linear in size of the domain ($k$).

\begin{figure}[ ]
\centering
\includegraphics[height=.26\textheight,width=\textwidth,keepaspectratio,trim=1cm 1cm 1cm 1cm]{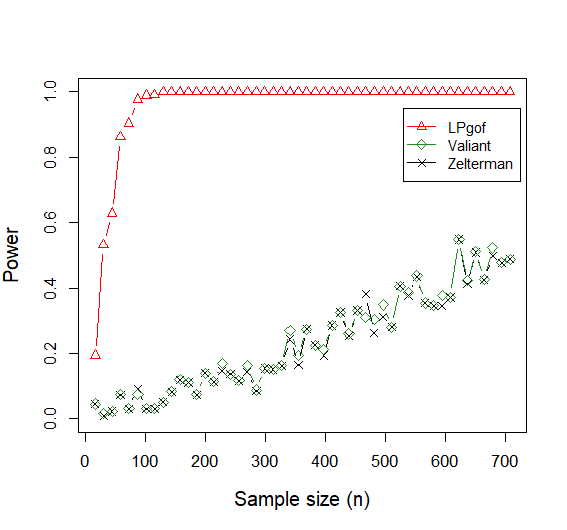}~~~~~~~~
\includegraphics[height=.26\textheight,width=\textwidth,keepaspectratio,trim=1cm 1cm 1cm 1cm]{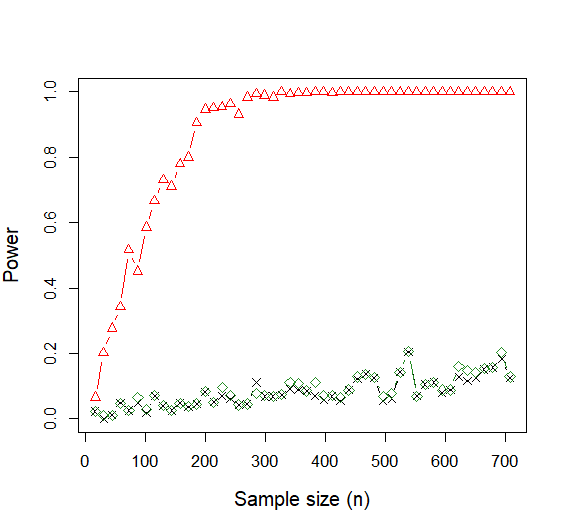} \\[1em]
\includegraphics[height=.26\textheight,width=\textwidth,keepaspectratio,trim=1cm 1cm 1cm 1cm]{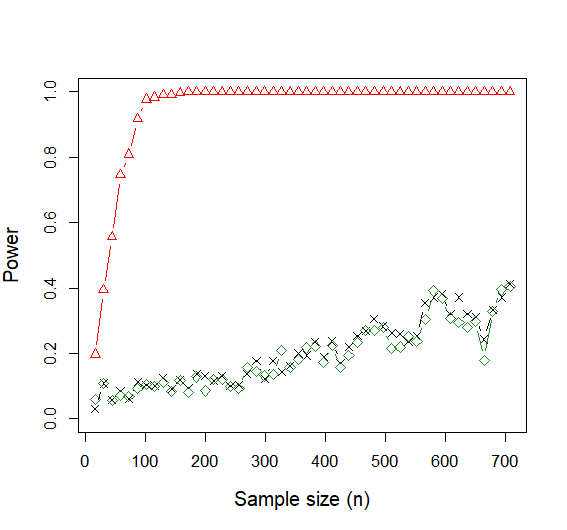}~~~~~~~~
\includegraphics[height=.26\textheight,width=\textwidth,keepaspectratio,trim=1cm 1cm 1cm 1cm]{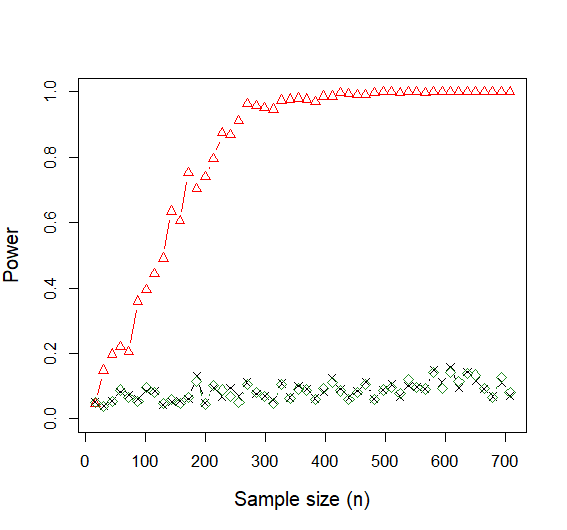} \\[1em]
\includegraphics[height=.26\textheight,width=\textwidth,keepaspectratio,trim=1cm 1cm 1cm 1cm]{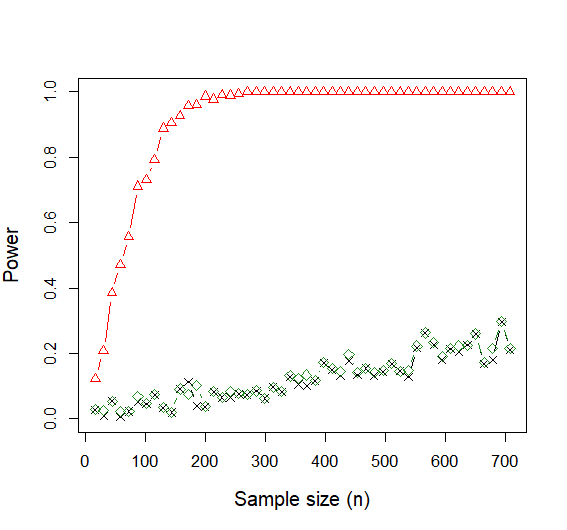}~~~~~~~~
\includegraphics[height=.26\textheight,width=\textwidth,keepaspectratio,trim=1cm 1cm 1cm 1cm]{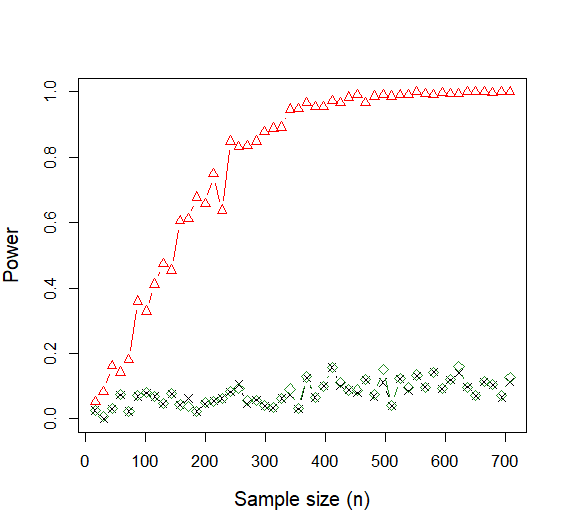} \\[1em]
\caption{First row: The null distribution is $p_0=U_{[k]}$ and step-function alternative $k^{-1}\al\, \ind_{1\le j \le k/2} + k^{-1}(2-\al)\, \ind_{k/2< j \le k}$ with $\al=.5, .7$. Second row: Construct discrete distribution as increments $\Delta F(j/n)$, $j=1,.\ldots,k$. For null model  we choose $p_0$ $={\rm Beta}(\al=2,\be=2)$, and for alternatives $p_\te$ is ${\rm Beta} (\al=2,\be=2+\te)$ with $\te=1,.5$. Third row:  Null distribution is $p_0=U_{[k]}$. The `bumpy' alternative distribution is constructed by taking the mixture of $U_{[k]}$ and increments of ${\rm Beta}(10,10)$ with mixing proportion $.3$ and $.2$. The red power curve denotes the \texttt{LPgof} method (see Sec \ref{sec:DEL}) with $m=8$.} \label{fig:power2}
\end{figure}

\vskip.6em
Some interesting new ideas and results recently appeared in the Theoretical Computer Science (TCS) literature under the banner of `Sub-linear Algorithms for Big data,' `Big data Property testing,' etc; see, for example, \cite{canonne15a,indyk2012a,valiant11a,RF2012a,diakonikolas2016a,orlitsky03a,orlitsky2015a,jiao2015minimaxa}.
They promise to give 20th-century pioneering statistical modeling ideas a 21st-century makeover with new ideas from TCS. They deserve the credit for asking the right question.  However, all of these methods are \emph{too specialized} and develops ``new tricks'' for each new related learning problem, which makes them practically inconvenient. This paper presents a more systematic and adaptable strategy, which starts with the following question (the holy grail of data analysis):
\vskip.25em
\emph{How to find an efficient coordinate system adapted to the given data sets sitting on a high-dimensional discrete domain to deliver time and sample efficient computation?}
\vskip.25em
Section \ref{sec:DEL} of the main paper,  performed an extensive numerical comparison of the power of different methods. Here we perform six further experiments. The detailed settings and results are summarized in  Fig. \ref{fig:power2}, which further re-establish the power of LP-domain statistical learning for detecting sparse structure in big discrete distributions.

\clearpage

\subsection{Additional Figures}
Some figures of the main article have been moved to the Appendix to save space.
\vskip2.5em

\begin{figure}[h]
  \centering
\includegraphics[width=.625\linewidth,keepaspectratio,trim=2cm 1cm 2cm 1cm]{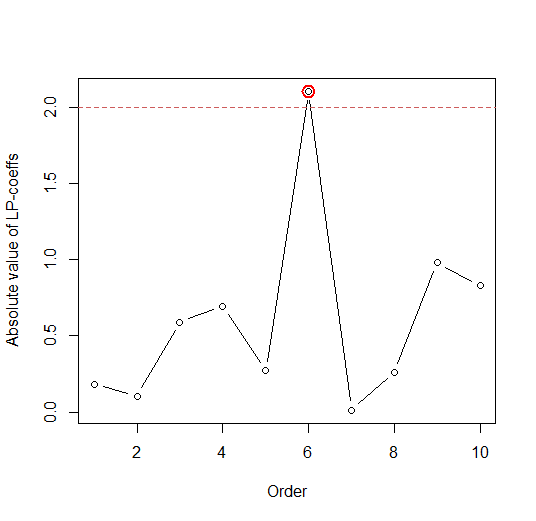}
\vskip1em
\caption{Earthquake Data: Plot of $\sqrt{n}\big| \tLP_j\big|$, $j=1,2,\ldots, 10$. The only significant component is $\sqrt{n}\tLP_6=2.10$, marked in red circle. see Sec. \ref{sec:dstheory} for more details.
.}\label{fig:earthlp}
\end{figure}

\begin{figure}[ ]
  \centering
\includegraphics[width=.45\linewidth,keepaspectratio,trim=1cm 1cm 1cm 1cm]{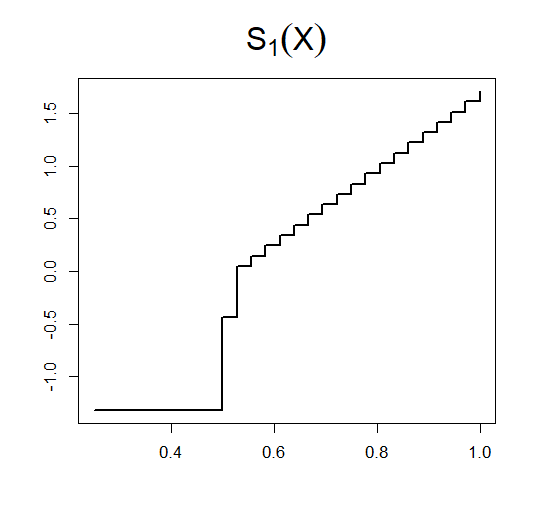}~~~~~
\includegraphics[width=.45\linewidth,keepaspectratio,trim=1cm 1cm 1cm 1cm]{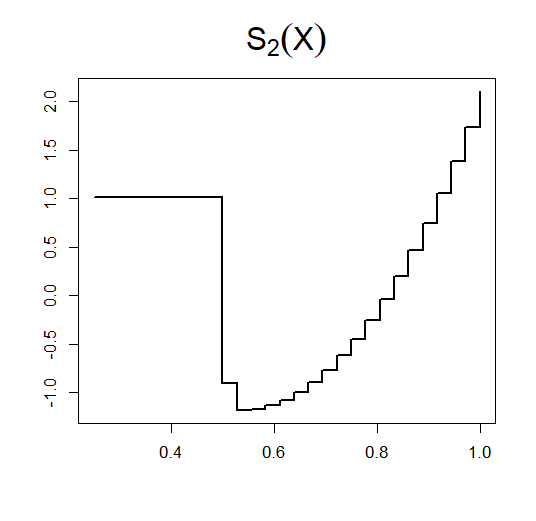}\\[2em]
\includegraphics[width=.45\linewidth,keepaspectratio,trim=1cm 1cm 1cm 1cm]{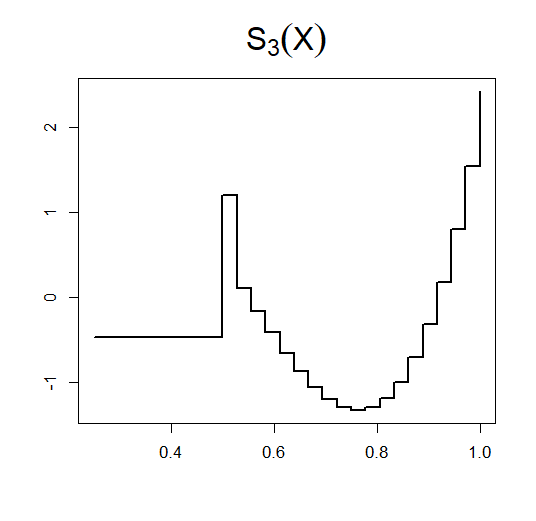}~~~~~
\includegraphics[width=.45\linewidth,keepaspectratio,trim=1cm 1cm 1cm 1cm]{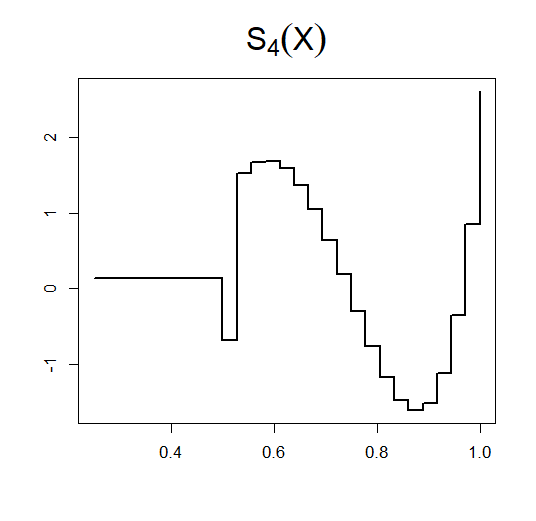}
\vskip1.5em
\caption{Sparse dice problem: Shape of the top four LP-orthonormal basis functions $\{S_j(u;F_0)\}_{1\le j \le 4}$ for $p_0(x)$ given in Example \ref{ex:SD}.}\label{fig:SD:basis}
\end{figure}

\begin{figure}[h]
  \centering
\includegraphics[width=.5\linewidth,keepaspectratio,trim=2.2cm 1cm 2.2cm 1.55cm]{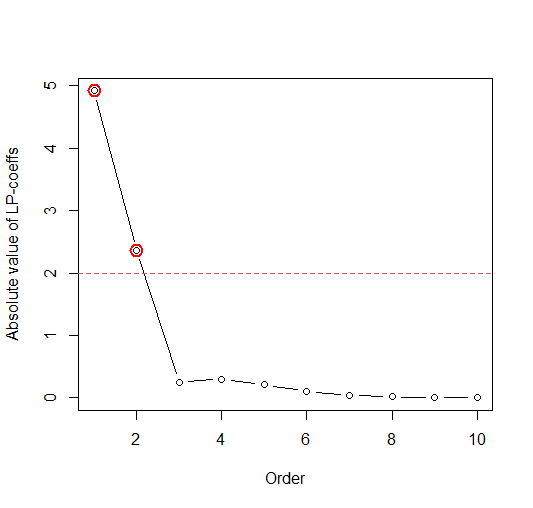}
\vskip1em
\caption{Sparse dice problem: Plot of $\sqrt{n}\big| \tLP_j\big|$, $j=1,2,\ldots, 10$. The two significant LP-Fourier coefficients are marked with red circles.}\label{fig:sdice}
\end{figure}

\begin{figure}[ ]
  \centering
\includegraphics[width=.464\linewidth,keepaspectratio,trim=1.15cm 1cm 1cm 1cm]{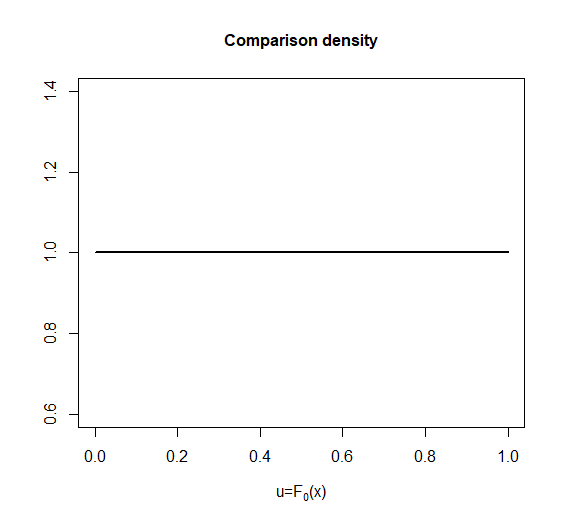}~~~~~~ \includegraphics[width=.464\linewidth,keepaspectratio,trim=1cm 1cm 1.15cm 1cm]{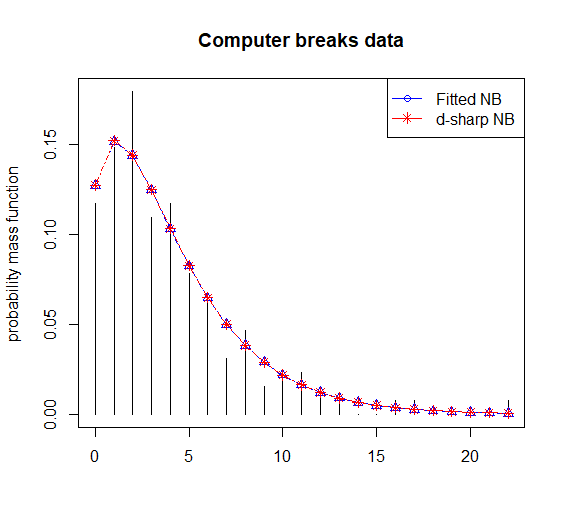}
\vskip.3em
\caption{Shows the analysis of $X$, which denotes a discrete random variable = number of times computer broke down in each of 128 consecutive weeks of operation. See Example \ref{ex:comp} of Section \ref{sec:dnb} for more details. The flat comparison density indicates that the negative binomial (with $\hat \mu =4$ and $\hat \phi=1.70$) adequately captures the pattern in the data. No further adjustments are needed.}\label{fig:comp}
\end{figure}

\begin{figure}[ ]
  \centering
\includegraphics[width=.55\linewidth,keepaspectratio,trim=2cm 1cm 2cm 1cm]{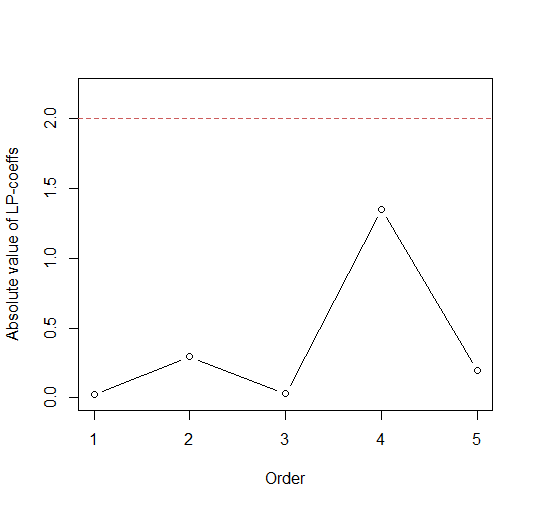}
\vskip1em
\caption{Spiegel Family Data: Plot of $\sqrt{n}\big| \tLP_j\big|$, $j=1,2,\ldots, 5$ where the $p_0(x)$ is taken to be ${\rm Binomial}(5, p=0.4625)$. None of the LP-coefficients are significant, thus the Binomial-model is accepted. see Example \ref{ex:spiegel} for more details. 
.}\label{fig:spiegel}
\end{figure}

\begin{figure}[ ]
  \centering
\includegraphics[width=.55\linewidth,keepaspectratio,trim=2cm 1cm 2cm 1cm]{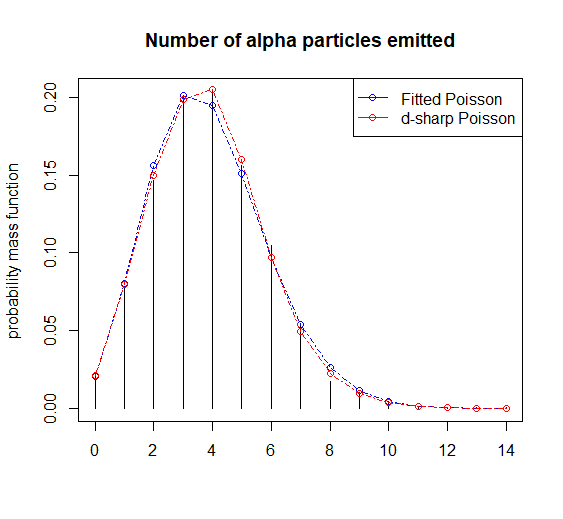}
\vskip1em
\caption{Rutherford-Geiger Polonium Data: The estimated ${\rm DS}(p_0,m)$ model is shown in red. Here $p_0(x)$ is ${\rm Poisson}(\la=3.87)$ shown in blue. see Example \ref{ex:polo} for more details.}\label{fig:polo}
\end{figure}

\begin{figure}[ ]
  \centering
\includegraphics[width=.64\linewidth,keepaspectratio,trim=2cm 1cm 2cm 1cm]{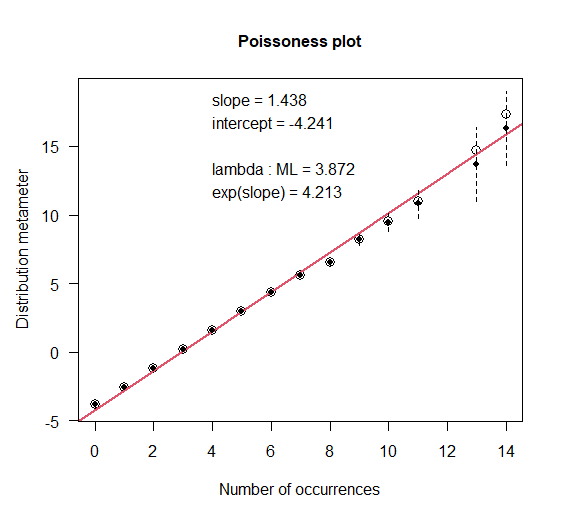}
\vskip2em
\caption{Rutherford's polonium data. Hoaglin's (1980) diagnostic plot to measure the ``Poissonness'' of the data. Note that when the data truly generated from  ${\rm Poisson}(\la)$ we would have $\widetilde{\phi}(x):=\log(x!\tp(x))$ will be close to a linear function of $x$, given by $\la +(\log \la) x$. The empirical function $\widetilde{\phi}(x)$ is called `\texttt{metameter},' denoted by open circles in the above plot. The dashed lines denote the confidence intervals, computed following \citet[Ch. 9]{hoaglin1985a}; centers of those intervals are marked with dark circles. The length of the CIs are large for data-sparse points $x\ge 11$; see Table \ref{tab:RFord}. The Poisson model is slightly dubious for the polonium data because the fitted line is outside the confidence interval at $x=8$. However, this is not a make-or-break factor; According to Tukey and Hoaglin ``we need to recall that one miss out of 15 is very common at 5\%.'' This perfectly matches with our assessments; see Example \ref{ex:polo} of Sec. \ref{sec:Xgof}.}\label{fig:poloeda}
\end{figure}

\begin{figure}[ ]
  \centering
\includegraphics[width=.55\linewidth,keepaspectratio,trim=2cm 1cm 2cm 1cm]{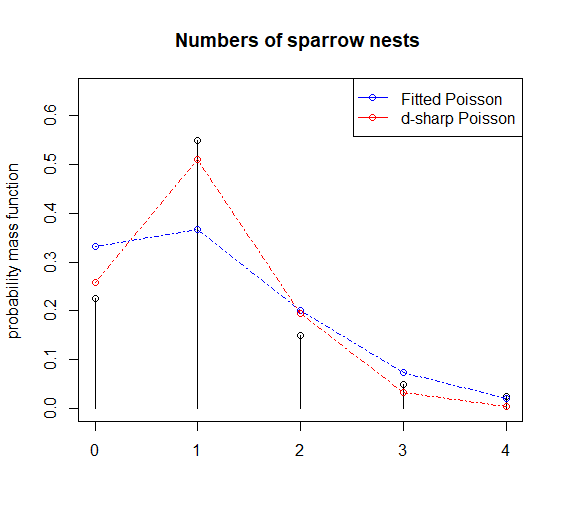}
\vskip1em
\caption{Sparrow data: $\DS(p_0,m)$ finds a under-dispersed Poisson model, which has a distinct peak at $x=1$.}\label{fig:sparrow}
\end{figure}

\begin{figure}[ ]
  \centering
\includegraphics[width=.5\linewidth,keepaspectratio,trim=2.25cm 1cm 2.25cm 1cm]{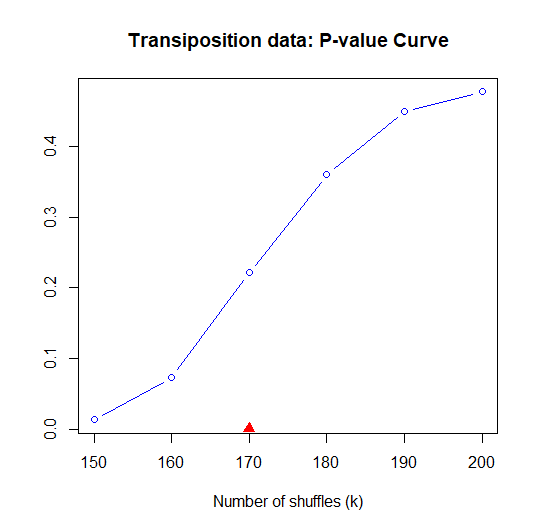}
\vskip1em
\caption{Card shuffling data (Sec. \ref{sec:card}):  $P$-value for \texttt{CARD}($k,n$) data with $n=500$ as a function of number of shuffles $k$. At each $(n,k)$ combination, we generated $B=250$ sets of random sample of size $n=500$. The average of those $B$  $p$-values are displayed, which shows $k=170$ random transpositions may be enough to produce a well-mixed deck.}\label{fig:card2}
\end{figure}

\begin{figure}[ ]
\centering
\includegraphics[width=.5\linewidth,keepaspectratio,trim=2.25cm 1cm 2.25cm 2.5cm]{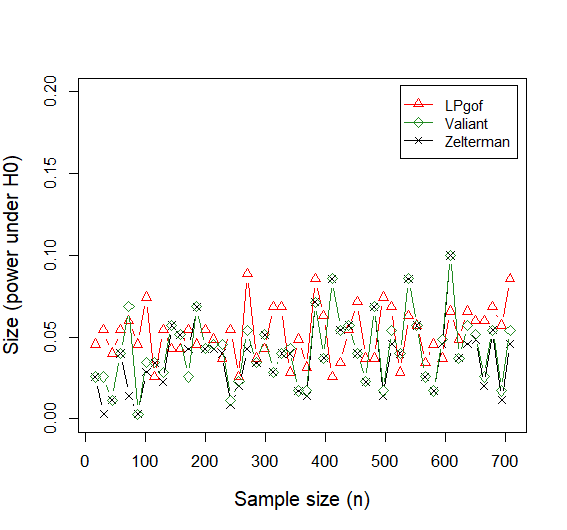}
\vskip1em
\caption{To examine whether the type-I error is controlled, we simulated data under the null $p_0=U_{[k]}$ for $k=5000$ with sample size $n$ ranging from $15$ to $700$. All the methods maintain type-I error rate quite well.}\label{fig:powerh0}
\end{figure}

\end{document}